\documentclass[%
 aip,
 jmp,%
 amsmath,amssymb,
reprint,%
onecolumn]{revtex4-1}

\usepackage{graphicx}
\usepackage{dcolumn}
\usepackage{bm}
\usepackage{hyperref}
\usepackage{natbib}
\usepackage{amssymb}
\usepackage{amsmath}
\usepackage{stackengine}
\usepackage{pgfplots}
\usepackage{adjustbox}
\usepackage{makecell, multirow, tabularx}

\usepackage{float}
\usepackage{caption}
\usepackage{subcaption}
\pgfplotsset{compat=1.18}

\begin{document}

\title{High-Frequency Stock Market Order Transitions during the US-China Trade War 2018: A Discrete-Time Markov Chain Analysis}

\author{Salam Rabindrajit Luwang}
\email{salamrabindrajit@gmail.com}
\affiliation{Department of Physics, National Institute of Technology Sikkim, Sikkim, India-737139.}

\author{Anish Rai}
\email{anishrai412@gmail.com}
\affiliation{Department of Physics, National Institute of Technology Sikkim, Sikkim, India-737139.}

\author{Md.Nurujjaman}
\email{md.nurujjaman@nitsikkim.ac.in}
\affiliation{Department of Physics, National Institute of Technology Sikkim, Sikkim, India-737139.}
	
\author{Om Prakash}
\email{omp1985@gmail.com}
\affiliation{Department of Mathematics, National Institute of Technology Sikkim, Sikkim, India-737139.}

\author{Chittaranjan Hens}
\email{chittaranjanhens@gmail.com}
\affiliation{Center for Computational Natural Sciences and Bioinformatics, International Institute of Information Technology, Hyderabad, India-500032.}

\date{\today}

\begin{abstract}
Statistical analysis of high-frequency stock market order transaction data is conducted to understand order transition dynamics. We employ a first-order time-homogeneous discrete-time Markov chain model to the sequence of orders of stocks belonging to six different sectors during the USA-China trade war of 2018. The Markov property of the order sequence is validated by the Chi-square test. We estimate the transition probability matrix of the sequence using maximum likelihood estimation. From the heat-map of these matrices, we found the presence of active participation by different types of traders during high volatility days. On such days, these traders place limit orders primarily with the intention of deleting the majority of them to influence the market. These findings are supported by high stationary distribution and low mean recurrence values of add and delete orders. Further, we found similar spectral gap and entropy rate values, which indicates that similar trading strategies are employed on both high and low volatility days during the trade war. Among all the sectors considered in this study, we observe that there is a recurring pattern of full execution orders in Finance \& Banking sector. This shows that the banking stocks are resilient during the trade war. Hence, this study may be useful in understanding stock market order dynamics and devise trading strategies accordingly on high and low volatility days during extreme macroeconomic events. 

\end{abstract}

\maketitle

\begin{quotation}
The stock market serves as a complex dynamical system, shaped by diverse traders and investors who employ various types of limit orders. During extreme market conditions, understanding the order dynamics is important to extract hidden information. Here, we analyse the high-frequency stock market order transactions during the US-China Trade war of 2018 to understand order transition dynamics and provide us with the ability to make rational choices.  We apply Markov property to a sequence of order data and estimate the transition probabilities. The Chi-square test validates the application of Markov property.  Based on these probabilities, we calculate the stationary distribution, mean recurrence time, spectral gap, relaxation rate and entropy rate.  An active participation of different kinds of traders was found on high volatility days.  Traders also place limit orders primarily with the intention to delete it. This shows that during extreme market situations, a lot of fake orders are placed to influence the market. Notably, the Finance \& Banking sector demonstrates resilience amid the trade war's uncertainty. The findings from this analysis hold the potential to guide intraday traders and investors in making informed and smart decisions.
\end{quotation}

\section{Introduction}
\label{sec:sample1}

In recent years, the study of stock market dynamics has gained significance in order to understand the trading behaviour of the market participants and the overall characteristics of the market in general ~\cite{le2020multiple,ricco2022information,mahata2020identification,mahata2020time,mahata2021characteristics}. In an electronic order-driven market, anonymous traders place orders in the limit order book. After a limit order has been placed, the order may be fully executed, partially executed, revised, or cancelled. It has been observed that a significant number of limit orders are amended or cancelled following their submission~\cite{fong2010limit}. With the advancement in trading technology and increase in active high-frequency trading, not only the number but also the frequency of order revisions and cancellations has also increased rapidly~\cite{le2020multiple, kirilenko2017flash, hasbrouck2013low}. Market latency has been improved to $300$ microseconds and the order capacity has increased to $100,000$ orders per second, as early as in $2010$~\cite{le2020multiple}. The orders-to-trades ratio, which indicates the intensity of order cancellations, have also risen significantly following a structural change towards market automation~\cite{hendershott2011automation}. Hence, understanding stock market order transition dynamics is essential for traders, investors and other market participants to manage risk and make informed decisions.

Market participants place different types of stock market orders with varying frequencies depending on the condition of the market and volatility of the day~\cite{le2020multiple,hagstromer2013diversity,kyrolainen2008day}. Volatility is the tendency for prices to unexpectedly change~\cite{harris2003trading} due to unexpected events. It is crucial to conduct thorough research on stock market order transition dynamics on high and low volatility days, especially during periods of economic uncertainty caused by significant macroeconomic events. A day is classified as highly volatile when the price difference between the highest and lowest price of the day is at its largest, and as lowly volatile when it is at its least. Notable examples of macroeconomic events include the financial crisis of 2008, the COVID-19 pandemic, and the USA-China trade war of 2018-2019, as they have the potential to induce high levels of volatility in the stock market~\cite{mahata2021modeling,rai2022statistical,rai2022sentiment,rai2023detection, baek2020covid, uddin2021effect}. In particular, the USA-China trade war, which commenced in early 2018 with mutually imposed tariffs, generated market volatility across various sectors of the economy, leading to fear and panic among investors~\cite{si2021policy}. During these periods of heightened volatility, the Chicago Board Options Exchange's Volatility Index (CBOE VIX), also known as the fear index, reached its highest levels since 2011 due to a substantial sell-off in the stock markets~\cite{VIX}. The increased volatility and a turbulent market can make it difficult for market participants to make informed decisions. Hence, this study aims to comprehensively analyze the dynamics of stock market order transitions on high and low volatility days during the USA-China trade war across six sectors in the USA: Energy, Finance \& Banking, FMCG, Healthcare, IT, and Real Estate. We have chosen these sectors as these sectors were impacted by the trade war~\cite{shi2021does, goulard2020impact,antoine2018costs}. The findings of this study can provide valuable insights for investors and traders in making informed decisions regarding market trends and risk management during extreme macroeconomic events.

Numerous research studies have explored the impact of the US-China trade war on the stock market. For example, researchers have used the Diebold-Yilmaz framework and a complex network approach to examine spillover effects between foreign capital in China's stock market and other financial assets. Their findings indicate a significant connection between China's stock market and international markets, with money flows being influenced by global instability, including the Sino-US trade war~\cite{xu2023impacts}. Researchers have also conducted a comprehensive analysis to assess the impact of investor sentiment on stock market returns and volatility connectedness during the trade war~\cite{bissoondoyal2022sentiment}. Time-varying stock market co-movement between the US and China was conducted at market level and sector level~\cite{shi2021does}. Furthermore, studies have explored the degree of dependence between the US stock market and stock markets in BRICV (Brazil, Russia, India, China, Vietnam) countries. The results have shown high unconditional volatility in stock returns and a dynamic correlation between the US stock market and each of the BRICV countries' stock markets~\cite{saijai2021measuring}. Moreover, Oh and Kim approached the effect of the trade war on stock markets from a financial contagion perspective using high-frequency financial data~\cite{oh2021effect}. However, no literature is found that deals with the study of the dynamics of stock market order transitions during the USA-China trade war.

In the stock market, different types of orders, as shown in Table~\ref{tab:EventDescription} of the Appendix, transition from one to another at an extremely rapid pace, even in microseconds~\cite{cont2011statistical}. A sequence of categorical data is formed by the transition between different types of orders. Analyzing the dynamics of these high-frequency categorical data sequences require different approaches, among which Markov-based models are commonly used for modeling such sequences~\cite{huang2001cube,ching2004higher,michelot2016movehmm,franke2004analysis,holzmann2006hidden,ching2002multivariate}. Markov model-based techniques have been extensively researched and applied in various fields; physics~\cite{van2018simple,fitzpatrick2020thermodynamics}, chemistry~\cite{anderson2011continuous,kutchukian2009fog}, biology and medicine~\cite{george2009towards}, social sciences~\cite{heneman1977markov,dijkstra2019networks} and queuing theory~\cite{meyn2008control}. In the field of finance and stock market too, Markov-based models have been used.

Markov-based models have been employed to investigate various phenomena in finance and the stock market, including market dynamics and transitions~\cite{siu2005multivariate, jones2005estimating, thyagarajan2005retail}. For instance, a semi-Markov approach was used to study high-frequency price dynamics. This approach models returns by assuming that intraday returns follow a discrete-time homogeneous semi-Markov process. The model incorporates a memory index to capture periods of high and low volatility in the Italian stock market~\cite{d2012semi, d2011semi}. Additionally, a generalized regime-switching model of the short-term interest rate was developed, where a first-order Markov process with state-dependent transition probabilities governs the switching between regimes~\cite{gray1996modeling}. Government spending and taxes was modelled as time-varying transition probability Markovian processes to tests for nonlinear effects of asset prices on the US fiscal policy~\cite{agnello2013using}. Credibility theory was used to estimate credit transition matrices in a multivariate Markov chain model for credit rating~\cite{siu2005multivariate}. In addition to these Markov-based models, other approaches such as agent-based models~\cite{raddant2017transitions}, smooth transition regression models~\cite{granger1993modelling} and GARCH models~\cite{chou2012euro} have been used to understand the dynamics and transitions in stock market. In this study, we chose discrete time Markov chains (DTMC), a Markov based model due to its simplicity and efficiency~\cite{badis2015modeling}. We have not found any work on the analysis of transition dynamics of different stock market orders using Markov models. Hence, a detailed study on the analysis of order transition dynamics during extreme market conditions using high-frequency stock market order data is very important.

Markov chains are classified as: continuous time Markov chains (CTMC) and discrete time Markov chains (DTMC). In DTMC, the state changes only at the discrete instants whereas the state can change at any time for CTMC. Mathematically, a DTMC is a sequence of random variables called states $X_n, \text{where}~ n = 1, 2, 3, \dots$, with the Markov property which states that evolution of the Markov process in the future depends only on the present state and not on past history. The state space, $S$ is the set of all possible states. The set $S$ can be finite or countably infinite. The transition rule between states is specified by transition probabilities~\cite{mo2010performance}. These transition probabilities are said to be time-homogeneous if it is independent of the time parameter. In the case of high-frequency stock market order transitions where the states, i.e., orders are placed at discrete instants of time, DTMC is preferred over CTMC. The different types of orders which constitutes the state space $S$ is also finite. Overall, DTMC provides a framework for modeling the probabilistic relationships between successive orders in the sequence, allowing us to understand and analyze different orders and their inter-dependencies. Further, the first-order DTMC model are relatively simple to implement and offer a clear and intuitive understanding of the order dynamics~\cite{ching2003higher,liu2003video,anderson1982holdings}. The first-order chains are preferred due to their minimal parameter estimation requirements and their usefulness in calculating other quantities~\cite{conejo2001first,coe1982fitting}. 

In this present work, we have used the first-order time-homogeneous DTMC with finite states to analyse the high-frequency stock market order transitions during the USA-China trade war of $2018$. The aim is to analyze the dynamics of order transition during high and low volatility days for six sectors. There are a total of ten possible stock market orders; $AB, AA, DB, DA, FB, FA, EB, EA, CB$ and $CA$,  which is described in Table~\ref{tab:EventDescription} of the Appendix. The orders are placed sequentially starting from the market opening at 4:00:00.000 EST until the market closes at 20:00:00.000 EST. In this process, a sequence of orders is formed where different orders transition from one to another. The probability associated with such a transition is referred to as the transition probability. These transition probabilities are estimated by using the Maximum Likelihood Estimation (MLE) Method. They are crucial for analyzing order transition dynamics and determining stationary distribution, mean recurrence time, spectral gap, relaxation rate, and entropy rate. PyDTMC, a Python package, is employed for these calculations. The findings may offer insights into the order dynamics during the USA-China trade in 2018.

The rest of the paper is organised as follows: Section~\ref{sec:DD} describes the high-frequency stock market order transaction data, Section~\ref{sec:M} describes the methods used for the analysis. The results are discussed in Section~\ref{sec:R}. Finally, Section~\ref{sec:Conc} gives the conclusion of our study.

\section{Data Description}
\label{sec:DD}
For this study, Algoseek~\cite{Algo} provided high-frequency stock market order transaction data. The data contained events that occurred during the entire trading day, from 04:00:00 EST to 20:00:00 EST. We have analysed the data from November 1st, 2018 to December 31st, 2018. The data was originally provided in the $BEFZ$ file format, which is unsuitable for analysis, and as a consequence, it was converted into $CSV$ files, which are more commonly used for data analysis. The converted $CSV$ data files were of considerably large size, ranging from $20$ to $40$ GB for one full day of trading data that included all listed stocks in NASDAQ. Such a large volume of data presented a significant challenge for data pre-processing. Therefore, we employed EmEditor, which is a powerful text editor that can handle large files and extract data efficiently.

The sample structure of the full data format is provided in Table~\ref{tab:DataSample} of the Appendix. The table contains eight columns, with the first column showing the date, the second column showing the timestamp, the third column showing the order ID, the fourth column showing the event type, the fifth column showing the ticker symbol of the traded stock, the sixth column showing the price at which the transaction occurred, the seventh column showing the number of stocks traded, and the eighth column showing the exchange on which the trade took place. Thus, the data contains information of all the types of orders placed starting from 04:00:00 EST to 20:00:00 EST. The length of this data varies for different days but they are in the range of 10s of millions. From this 10s of millions long data which is for all the listed stocks, we extract the data for 18 stocks, belonging to 6 sectors as shown in~\ref{tab:Stocks_Sectors}, for this study.

The event type in the fourth column of Table~\ref{tab:DataSample} are the different types of stock market orders submitted to NASDAQ. These orders will be referred to as states of the Markov chain. For this study, it is essential to understand the various types of orders submitted to the exchange as our focus is entirely on the sequence of these orders. Table~\ref{tab:EventDescription} in the Appendix presents the type of the order, the abbreviation used in the paper and their descriptions. The table outlines the BID and ASK orders linked with ADD, CANCEL, DELETE, EXECUTE and FILL. The BID order is used for purchasing, while the ASK order is used for selling. For example, if the order is ADD-BID, a buying order is placed in the exchange. To remove this order, either a CANCEL BID order or a DELETE BID order can be used. If the order is executed, it is indicated by either FILL BID or EXECUTE BID. A sample of the sequence of high-frequency stock market orders may be of the form: AB, AA, AB, DB, DA, EB, FA, AB, FB, CA, DA, AA, CB,.....and so on. The sequence, which is made up of ten types of orders, will be analyse by using a DTMC model.

\section{Methodology}
\label{sec:M}
In this section, we discussed the methodologies that are utilized for the purpose of analysing the sequence of high frequency stock orders. 
\subsection{Selection criteria of High and Low volatility day}
\label{subsec:DayCriteria}
To identify high and low volatility days, the following simple steps are undertaken:
\begin{enumerate}
\item[(i)]  We calculate the price difference between a stock’s highest and lowest values for a given day within the time-frame of 01-10-2018 to 31-12-2018.
\item[(ii)] We compute the mean ($\mu$) and standard deviation ($\sigma$) of the normalized price range (i.e., the highest minus the lowest price) for the stock.
\item[(iii)] Based on the obtained statistical measures, we consider the day as high volatility day if the price difference exceeds  the threshold of $(\mu+\sigma)$.
\item[(iv)] Similarly, we consider the day as low volatility day if the price difference is below the threshold of $(\mu-\sigma)$.
\end{enumerate}
By following these steps, we can effectively identify and categorize the level of volatility experienced on specific trading days.

\subsection{Dependency Test}
To examine the appropriateness of employing a Markov chain comprising $r$ states for analyzing a sequence of high-frequency stock market order transaction data, one can explore this by utilizing the Chi-square ($\chi^2$) test of independence. This entails assessing the independence or dependence of successive orders within the sequence.
The $\chi^2$-test statistic is given by:
\begin{equation}
    \chi^{2} = \sum_{i=1}^{r}\sum_{j=1}^{r} \frac{{(n_{ij}-e_{ij})}^2}{e_{ij}},
\end{equation}
where $n_{ij}$ represents the observed transition frequency, and $e_{ij}$ represents the expected transition frequency\cite{holmes2021discrete}. The observed transition frequency is the actual number of transitions between different states from the data whereas the expected transition frequency is the number of transitions we would expect to occur if there were no relationship or association between the states. $e_{ij}$ is calculated based on the assumption of independence between the states. 

The null hypothesis of the test is that the observed data are serially independent. The alternative hypothesis is that the observed data were generated by a Markov chain. Under the null hypothesis of independence, $e_{ij}$ is calculated as follows:
\begin{equation}
    e_{ij}=\frac{(\sum_{i=1}^{r}n_{ij})(\sum_{j=1}^{r}n_{ij})}{(\sum_{i=1}^{r}\sum_{j=1}^{r}n_{ij})}
\end{equation}
For a Markov chain with $r$ states, the test statistics follows the $\chi^2$ distribution with ${(r-1)}^2$ degrees of freedom.

\subsection{Markov Chain}
Markov chain belongs to a category of stochastic processes that prove valuable in describing a sequence of categorical events. They are particularly useful when the likelihood of an event relies solely on the state of preceding events~\cite{de2019markov}. Such stochastic processes can be broadly categorized into four types: discrete-time Markov chain with discrete states, discrete-time Markov chain with continuous states, continuous-time Markov chain with discrete states, and continuous-time Markov chain with continuous states~\cite{gao2020markov}. In this paper, as the state space, $S$ is finite and the state transitions are assumed to be happening at discrete intervals of time, we will focus on a discrete-time Markov chain with discrete states. 

\subsubsection{Discrete-Time Markov Chain}
A Discrete-Time Markov Chain (DTMC) refers to a series of random variables denoted as $X_1, X_2, \ldots, X_n$, which follows the Markov property~\cite{spedicato2016markovchain}. The Markov property states that the transition probability from one state to another in the sequence depends solely on a finite set of preceding states. To establish an accurate model, it is essential to determine the number of preceding states, known as the order of the Markov model. This study focuses solely on the first order, where the transition probabilities for future states can be determined based only on the current state~\cite{shamshad2005first,conejo2001first,tang2020markov}. By adopting this simplifying assumption, the model becomes straightforward to implement. 

Mathematically, in the context of a first-order DTMC, the future state denoted as $X_{n+1}$ relies exclusively on the current state, represented as $X_n$~\cite{spedicato2016markovchain,balzter2000markov}, i.e.,

\begin{align}
    & P(X_{n+1} = x_{n+1}\mid X_1 = x_1, X_2 = x_2,....,X_n = x_n) \nonumber\\
    & P(X_{n+1} = x_{n+1}\mid X_n = x_n)
\end{align}
  
For this study, we have considered a DTMC with discrete and finite states. The set of possible states is $S = \{s_1,s_2,....,s_r\}$, where $r$ is the number of states, which is discrete and finite. The chain moves from one state to another state. The probability to move from state $s_i$ to state $s_j$ in one step (or transition) is called transition probability, denoted by $p_{ij}$~\cite{balzter2000markov}.
\begin{equation}
    p_{ij} = P(X_1 = s_j\mid X_0 = s_i)
\end{equation}

We also assume that there is no time dependence on the probability of state change. It implies that the Markov chain is time-homogeneous~\cite{masseran2015markov,ghosh2017application}. So,
\begin{align}
    p_{ij} & = P(X_{n+1} = s_j\mid X_n = s_i) \nonumber\\
           & = P(X_n = s_j\mid X_{n-1} = s_i) \nonumber\\
           & = .............................. \nonumber\\
           & = P(X_1 = s_j\mid X_0 = s_i)
\end{align}

The probability distribution of transitions from one state to another can be represented by
a transition matrix $P = {(p_{ij})}_{i,j}$, where each element of position $(i, j)$ represents the transition probability $p_{ij}$. For $r$ states, $P$ has the form~\cite{shamshad2005first,tang2020markov,balzter2000markov}
\begin{equation}
P = 
\begin{pmatrix}
p_{1,1} & p_{1,2} & \cdots & p_{1,r} \\
p_{2,1} & p_{2,2} & \cdots & p_{2,r} \\
\vdots  & \vdots  & \ddots & \vdots  \\
p_{r,1} & p_{r,2} & \cdots & p_{r,r} 
\end{pmatrix}
\end{equation}
subject to 
\begin{align}
    & 0\leq p_{ij}\leq 1, \forall i,j \in S \\
    & \sum_{j=1}^{r} p_{ij} = 1, \forall i \in S 
\end{align}

The Maximum Likelihood Estimation (MLE) method, as proposed by Anderson and Goodman~\cite{anderson1957statistical}, was employed to estimate each one-step transition probability, $p_{ij}$, within the transition probability matrix $P$. By utilizing this approach, the number of transitions from state $i$ to state $j$ in the data sequence represented as $n_{ij}$, was used to derive the maximum likelihood estimate of the one-step transition probability, as presented below~\cite{shamshad2005first,ghosh2017application,masseran2015markov}:
\begin{equation}
    p_{ij} = \frac{n_{ij}}{\sum_{j=1}^{r} n_{ij}}
\end{equation}

States $i$ and $j$ of a DTMC are considered to be communicating if they can be reached from each other~\cite{ching2006markov}. In the transition probability matrix $P$, if both $p_{ij} > 0$ and $p_{ji} > 0$, then these states are considered to be communicating. If all states in the Markov chain can communicate with each other, it is classified as irreducible, with a single communicating class. On the other hand, if a Markov chain has multiple communicating classes, it is referred to as reducible~\cite{ching2006markov}.

Further, a state is classified as recurrent if, once it is left, the chain is guaranteed to return to that state with a probability of one. However, if the probability of returning to that state is less than one, the state is considered transient~\cite{ching2006markov}. Defining mathematically,
\begin{equation}
    f_{ii} = P(X_n = i,\mid X_0 = i),
\end{equation}
for some $n\geq 1$. State $i$ will be recurrent if $f_{ii} = 1$ and transient if $f_{ii} < 1$. Additionally, a Markov chain is deemed aperiodic when all of its states are aperiodic. A state is considered aperiodic if it does not exhibit periodic behavior, whereas a state is classified as periodic if the chain can only return to that state at intervals that are multiples of a specific integer greater than one~\cite{ching2006markov}.

\subsubsection{Stationary Distribution}
The stationary distribution of a Markov chain refers to the limiting probability distribution that remains constant over time, representing the long-term probabilities. In the case of a finite-state Markov chain that is both ergodic and finite, each row of the limiting distribution ($\lim_{n\to\infty} P^n$) converges to the stationary distribution. A Markov chain is considered ergodic if it satisfies both the conditions of aperiodicity and irreducibility. Within this context, let $\pi_j$ denote the proportion of time spent in state $j$ in the long run. When a finite-state Markov chain is ergodic, it possesses a unique stationary distribution and $\pi_j$ can be calculated using the following equations~\cite{holmes2021discrete}:
\begin{align}
    & \pi_j = \sum_{i=1}^{k} \pi_{j}p_{ij} \\
    & \sum_{j=1}^{k} \pi_{j} = 1   
\end{align}

\subsubsection{Mean Recurrence Time}
Consider a system that begins with the state $s_{j}$. Let $f_{jj}^{(n)}$  be the probability that it returns to the state $s_{j}$ after $n$ transitions, where $n = 1, 2, 3,...$and so on.

The mean recurrence time of state $s_{j}$ is given by~\cite{heyman1995fundamental,medhi1994stochastic}
\begin{equation}
    \mu_{jj} = \sum_{n=1}^{\infty} nf_{jj}^{(n)}
\end{equation}

\subsubsection{Spectral Gap and Relaxation Rate}
Let $\{\lambda_i\}$ be the eigenvalues of $P$, with $|\lambda_1 | \geq |\lambda_2 | \geq.......\geq| \lambda_n |$. By Perron Frobenius's theory for nonnegative matrices~\cite{seneta2006non}, it can be concluded that $|\lambda_1| = 1$, and that $| \lambda_i | <1$ for all $2 \leq i \leq n$. Here, $P$ is irreducible and aperiodic. 

Let $\lambda^{*} = $max$\mid \lambda \mid :\lambda \neq 1$ be the second largest eigenvalue modulus (SLEM) of the transition probability matrix $P$. Then, $\gamma^{*} = 1 - \lambda^{*}$ is called the absolute spectral gap, and 
\begin{equation}
    \gamma = 1 - \lambda_2
\end{equation}    
is called the spectral gap of P. The relaxation time $t_{rel}$ is defined by

\begin{equation}
    t_{rel} = \frac{1}{\gamma^{*}}
\end{equation}

The rate at which the Markov chain converges to the uniform equilibrium distribution in the long run is governed by the SLEM. The smaller the SLEM, the faster the Markov chain converges to its equilibrium distribution.

\subsubsection{Entropy Rate}

The entropy rate of an ergodic Markov chain with a finite state space $S$ is a measure of the average amount of uncertainty in the next state of the chain, given the current state. A higher entropy rate indicates a less predictable system, whereas a lower entropy rate suggests a higher level of predictability. It is calculated as the sum of the entropies of the transition probabilities $p_{ij}$, weighted by the probability of occurrence of each state in the stationary distribution~\cite{cover1999elements,vegetabile2019estimating,ciuperca2005estimation}. We apply this definition to a stationary first-order time-homogeneous
Markov process, $\mathcal{X} = \{X_t\}$, $t\in \{0, 1, 2,....,n\}$ with finite-state
space $S = \{s_1, s_2,....,s_r\}$. The entropy rate of the process is as follows:
\begin{align}
     H(\mathcal{X}) & =  \lim_{n\to\infty} H(X_n\mid X_{n-1}, X_{n-2},....,X_0) \nonumber \\
    & = \lim_{n\to\infty} H(X_n\mid X_{n-1}) \nonumber\\
    & = \lim_{n\to\infty}\sum_{s_i \in S}^{} Pr(X_{n-1} = s_i) H(X_n\mid X_{n-1} = s_i) \nonumber\\
    & = - \lim_{n\to\infty}\sum_{s_i \in S}^{} Pr(X_{n-1} = s_i)\sum_{s_j \in S}^{} Pr(X_n = s_j\mid X_{n-1} = s_i)\log_2 Pr(X_n = s_j\mid X_{n-1}=s_i)     
\end{align}
The stationary assumption and an assumption of a time-homogeneous
Markov process imply $Pr(X_n = s_j\mid X_{n-1} = s_i) = p(s_i, s_j) = p_{ij}$. Noting
that $\lim_{n\to\infty} Pr(X_{n-1} = s_i) = \pi_i, \forall s_i\in S$, we can write 
\begin{equation}
    H(\mathcal{X}) = - \sum_{s_i, s_j\in S}^{} \pi_i p_{ij} \log_2 p_{ij}
\end{equation}

The entropy of a discrete random variable $X$, taking values from the state space $S$, varies between zero and log$_2 r$, where $r$ is the cardinality of $S$~\cite{vegetabile2019estimating}. 

\section{Results}
\label{sec:R}
In this section, we analyse the dynamics of high-frequency stock market order transitions during the trade war by utilizing the first-order time-homogeneous Discrete-Time Markov Chain (DTMC) model. The analysis is conducted for six sectors: Energy, Finance \& Banking, FMCG, Healthcare, IT, and Real Estate during high and low volatility days.

In Subsec.~\ref{subsec:Selectioncriteria}, we present the high and low volatility days for each stock. In Subsec.~\ref{subsec:DT}, we show the dependency between different types of orders in the order sequence as dependent using the Chi-square test of independence before considering the sequence as the chain of the model. Subsec.~\ref{subsec:TPM} describes the transition probability between the orders using the heatmap representation of the transition probability matrix. Subsec.~\ref{subsec:SD} describes the stationary distribution of each order, Subsec.~\ref{subsec:MFPT_MRT} gives the mean recurrence time of each order, Subsec.~\ref{subsec:SG_RR_ER} represents the spectral gap, relaxation rate and the entropy rate of the DTMC.

\subsection{Selection of High and Low volatility days}
\label{subsec:Selectioncriteria}

In this study, we have selected three active stocks listed in NASDAQ from each of the six sectors. The selection of the stocks was based on the largest market capitalization, as presented in Table~\ref{tab:Stocks_Sectors} of the Appendix. We identify a high volatility day and a low volatility day of a stock, following the methodology described in Section~\ref{subsec:DayCriteria}, to study the order dynamics during the trade war. The high and low volatility days of the selected stocks is shown in Table~\ref{tab:Stocks_Dates}.

\begin{table}[!h]
\begin{center}
\small
    \begin{tabular}{ | l | l | p{4.5cm} |}
    \hline
	\textbf{Date} & \textbf{High volatility stocks} & \textbf{Low volatility stocks} \\
	\hline
05-11-2018 & WFC & UNH, JPM, BAC  \\
\hline
06-11-2018 & - & XOM  \\
\hline
08-11-2018 & - & MSFT, PLD \\
\hline
09-11-2018 & - & MCD, AMT  \\
\hline
23-11-2018 & - & GOOGL  \\
\hline
26-11-2018 & - & JNJ, CVX, CCI  \\
\hline
29-11-2018 & - & CVS, PG, WFC  \\
\hline
06-12-2018 & UNH, CVS, PLD, MCK & - \\
\hline
13-12-2018 & - & NEE  \\
\hline
17-12-2018 & PG & - \\
\hline
21-12-2018 & GOOGL & -  \\
\hline
24-12-2018 & NEE, JNJ & - \\
\hline
26-12-2018 & CCI, AMT, BAC, JPM, MSFT,& \\
           &  AAPL, CVX, XOM, MCD & - \\
\hline
31-12-2018 & - &  AAPL, MCK\\
    \hline
    \end{tabular}
    \caption{High and Low volatility days for each stock.}
    \label{tab:Stocks_Dates}
\end{center}
\end{table}

\subsection{Dependency Test}
\label{subsec:DT}
To assess the suitability of applying DTMC to high-frequency stock market order sequence for all the stocks, we conducted a Chi-square ($\chi^{2}$) test of independence. For each order sequence, we calculated the $\chi^{2}$-Statistic and compared it to the $\chi^{2}$ distribution with $(r-1)^2$ degrees of freedom at a significance level of $1\%$.

Table~\ref{tab:Chi} in the Appendix presents the calculated values of $\chi^2$-Statistic and the corresponding $p$-values for each data sequence associated with the respective stocks. Notably, all the calculated $p$-values in the table are found to be very less than $0.01$. Therefore, based on the $1\%$ level of significance, we reject the null hypothesis which states that observed data are serially independent and accept the alternative hypothesis which states that the observed data were generated by a Markov chain. This leads us to conclude that the sequential data exhibit serial dependency, affirming the feasibility of applying the Markov property.

\subsection{Transition Probability Matrix of the DTMC Model}
\label{subsec:TPM}
The transition probabilities of the stock market orders, $AB,AA,CB,CA,DB,DA,EB,\\EA, FB$ and $FA$, are estimated using the MLE method.
These probabilities constitutes the transition probability matrix of the DTMC model. In total, there are twelve transition probability matrices for the six sectors. Two probability matrices belong to a sector; one for the average of high volatility days and another for the average of low volatility days. We will analyse these twelve matrices and compare the dynamics of order transitions on high and low volatility days during the trade war for each sector. 

\begin{figure}
\centering
\begin{subfigure}{0.375\textwidth}
    \includegraphics[width=\textwidth]{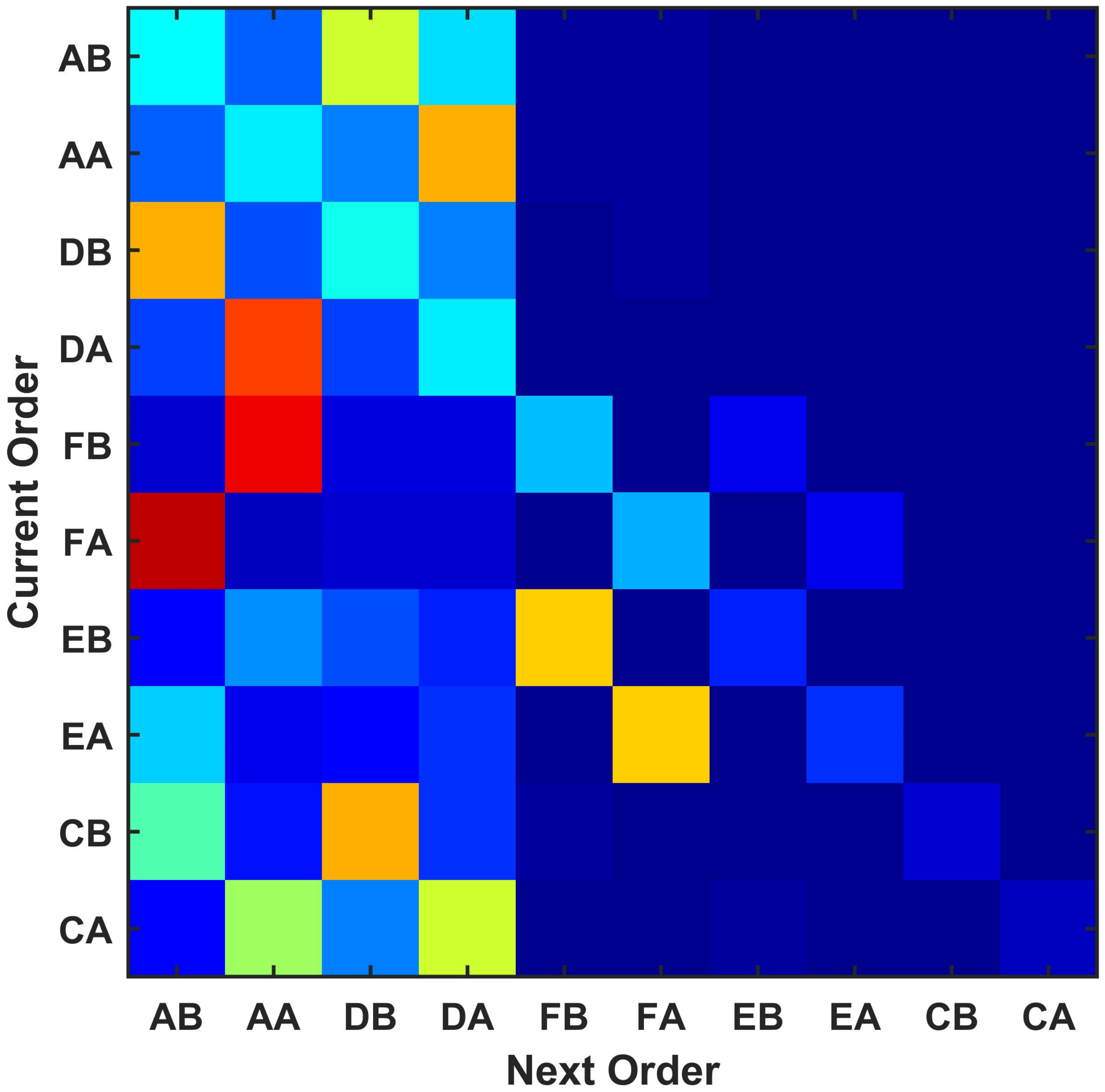}
    \caption{Healthcare (High volatility)}
    \label{subfig:H_H}
\end{subfigure}
\hspace{1cm}
\begin{subfigure}{0.405\textwidth}
    \includegraphics[width=\textwidth]{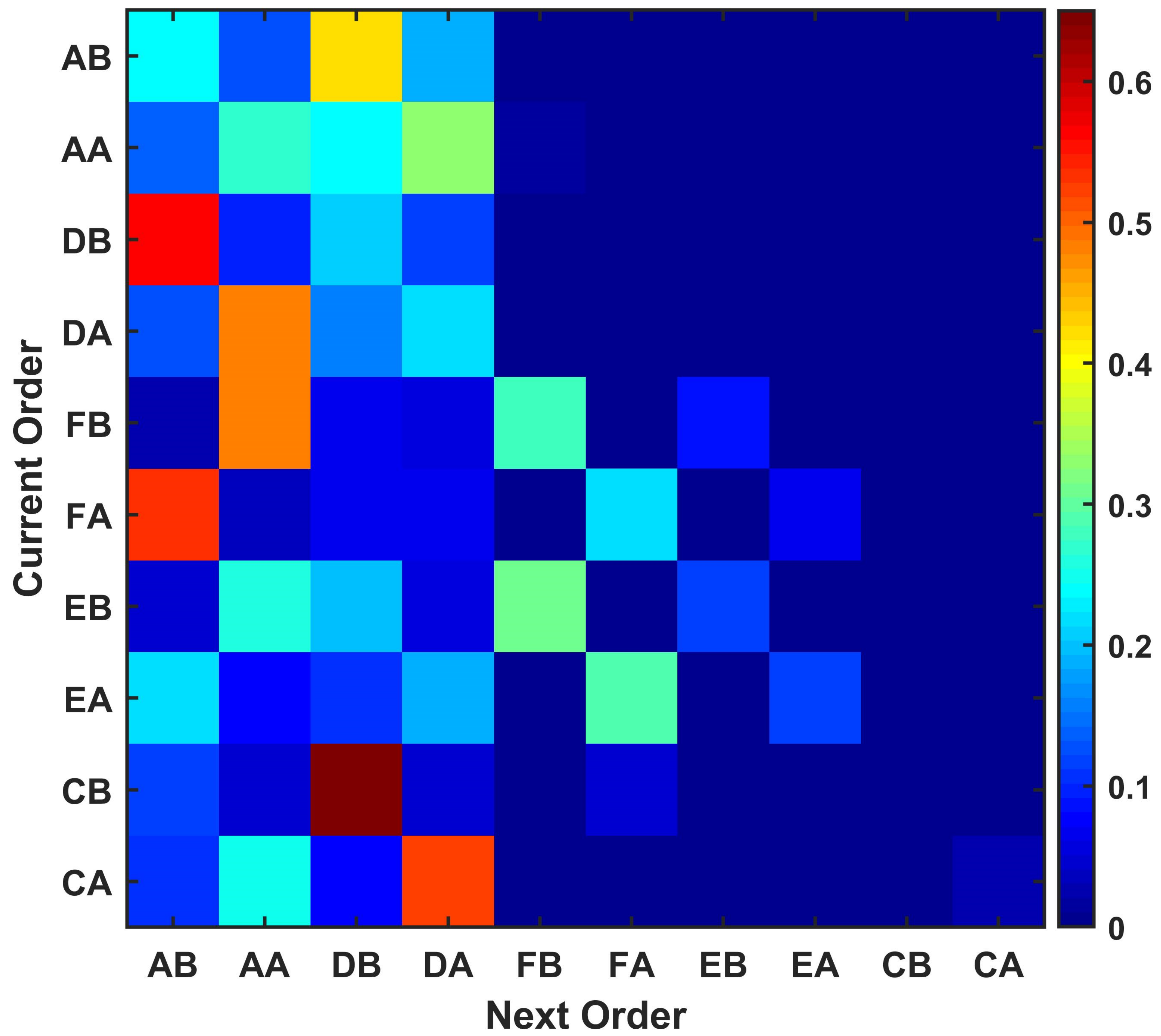}
    \caption{Healthcare (Low volatility)}
    \label{subfig:H_L}
\end{subfigure}
\begin{subfigure}{0.375\textwidth}
    \includegraphics[width=\textwidth]{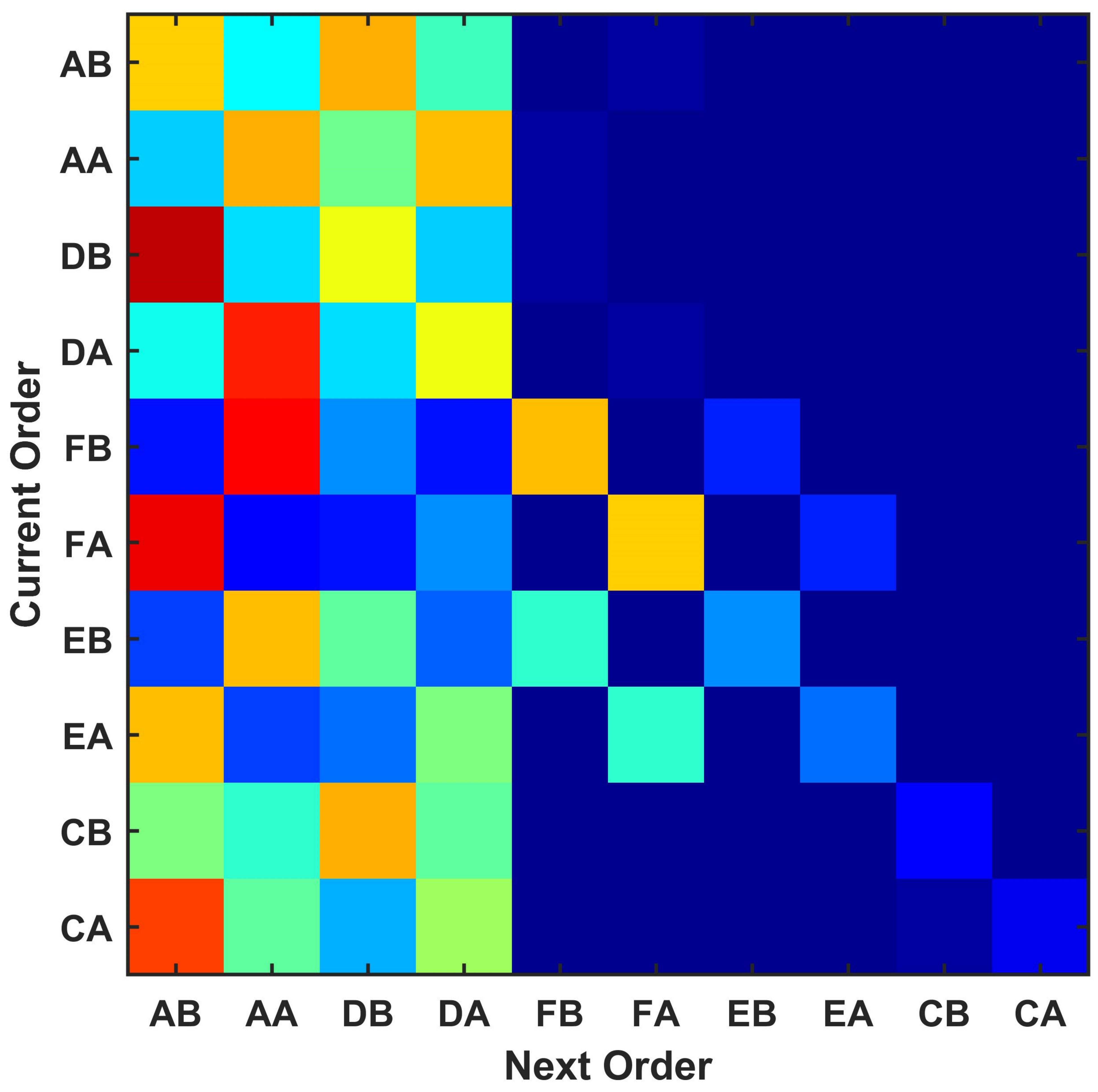}
    \caption{IT (High volatility)}
    \label{subfig:IT_H}
\end{subfigure}
\hspace{1cm}
\begin{subfigure}{0.405\textwidth}
    \includegraphics[width=\textwidth]{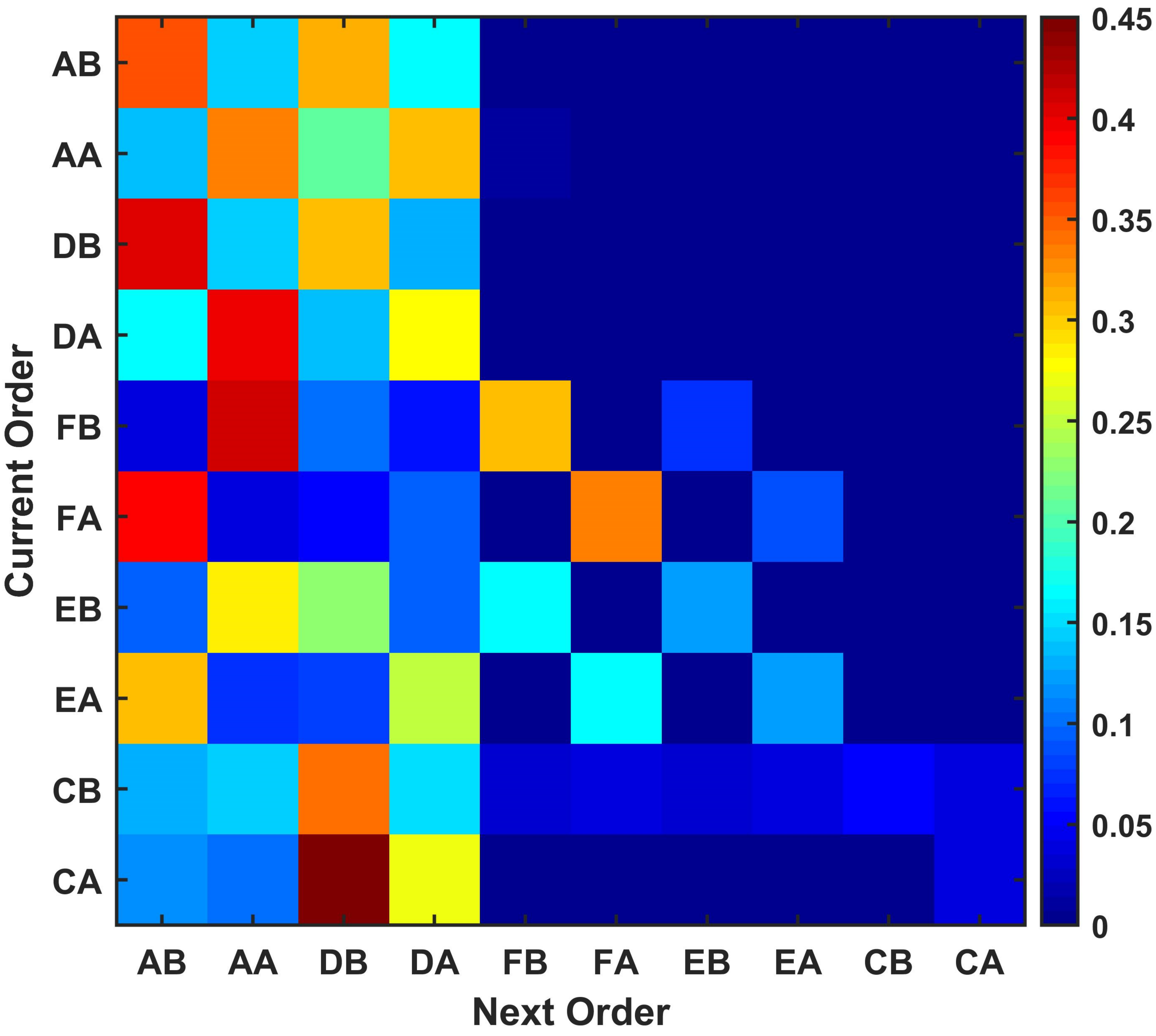}
    \caption{IT (Low volatility)}
    \label{subfig:IT_L}
\end{subfigure}
\begin{subfigure}{0.375\textwidth}
    \includegraphics[width=\textwidth]{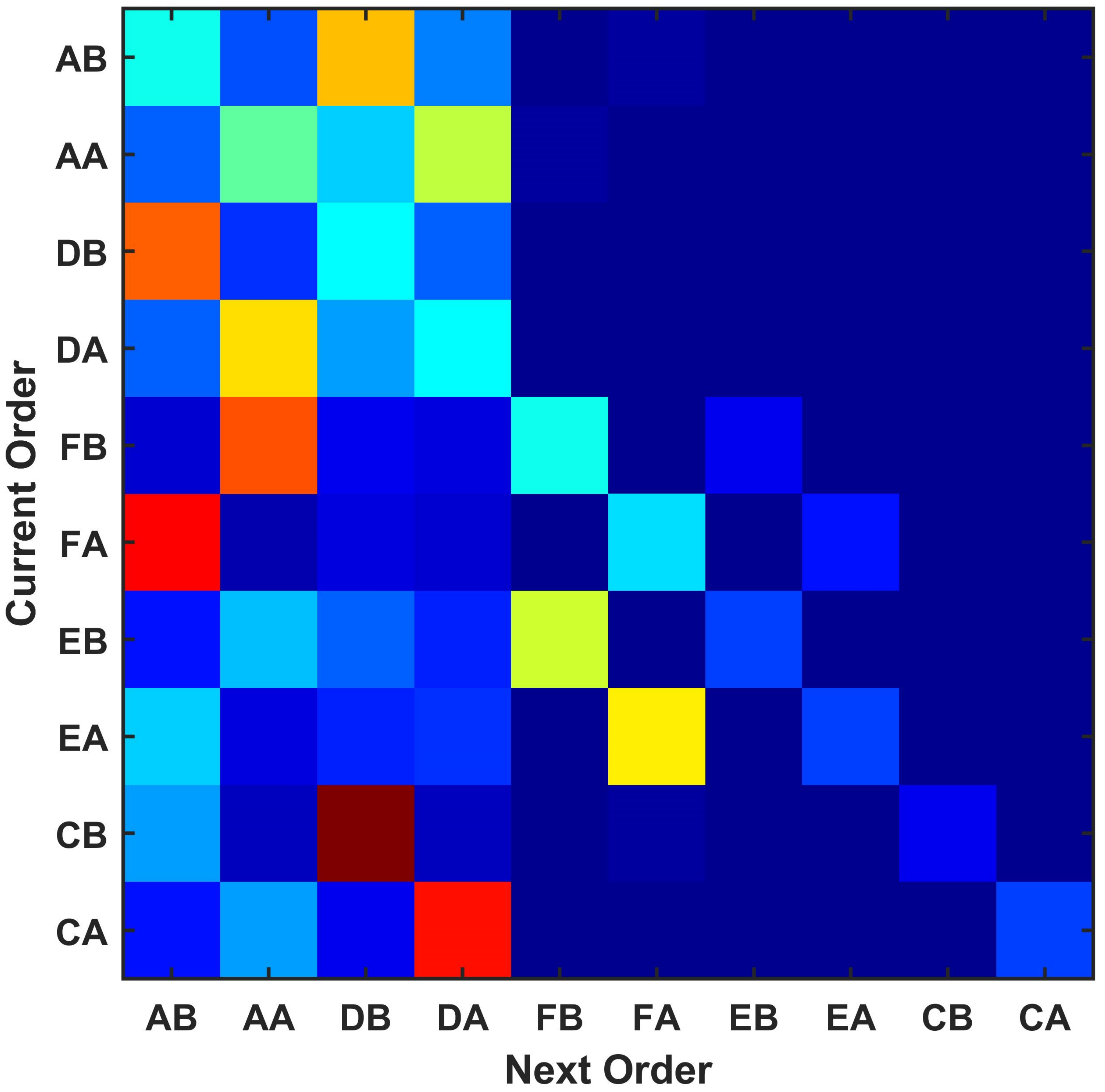}
    \caption{Real Estate (High volatility)}
    \label{subfig:RS_H}
\end{subfigure}
\hspace{1cm}
\begin{subfigure}{0.405\textwidth}
    \includegraphics[width=\textwidth]{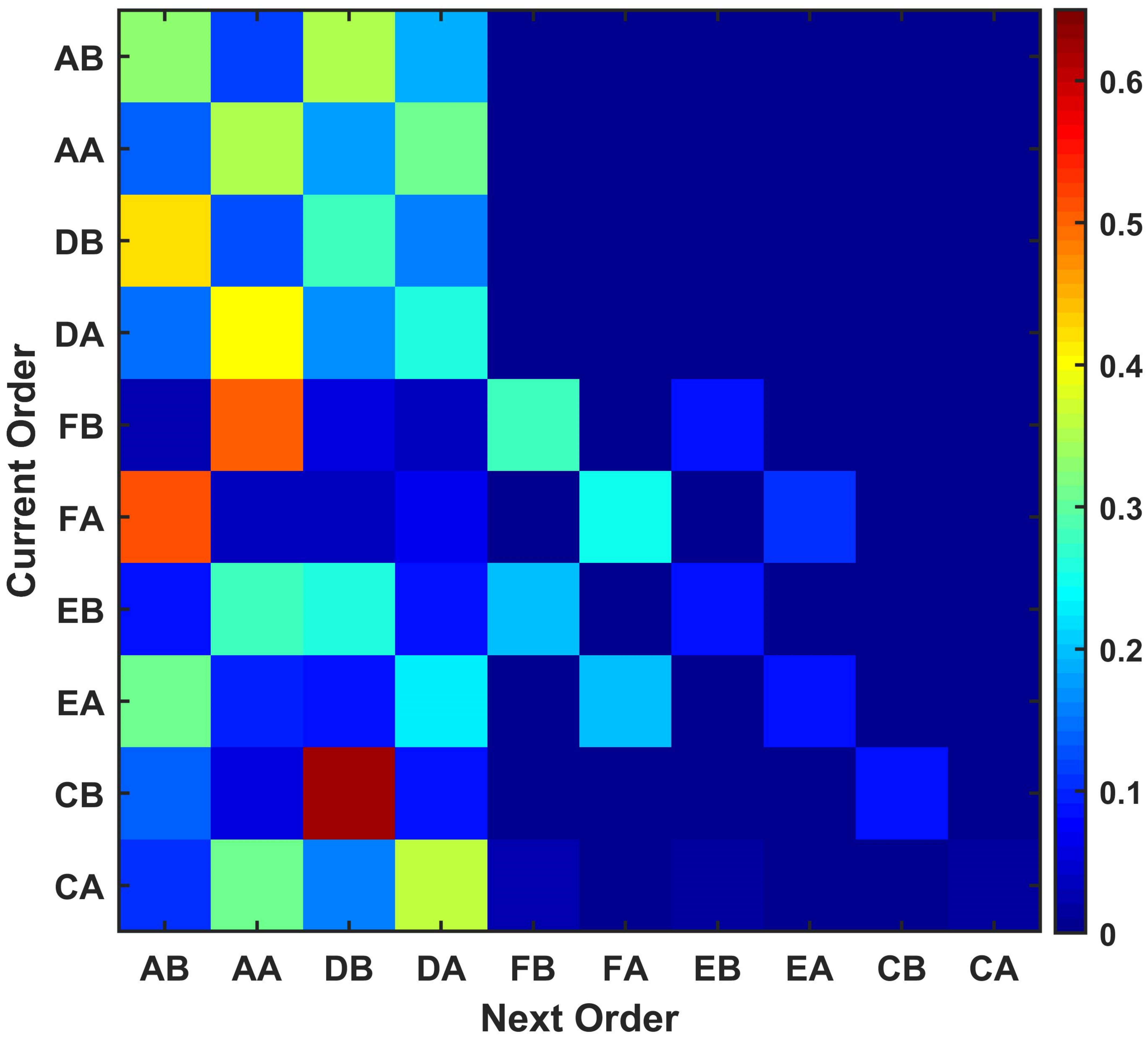}
    \caption{Real Estate (Low volatility)}
    \label{subfig:RS_L}
\end{subfigure}      
\caption{Heatmap representation of the transition probability matrix on high and low volatility days.}
\label{fig:Heatmap2}
\end{figure}

Figures~\ref{subfig:E_H} to~\ref{subfig:FMCG_L} of~\ref{fig:Heatmap1} and~\ref{subfig:H_H} to~\ref{subfig:RS_L} of~\ref{fig:Heatmap2} shows the heatmap comparison of transition probability matrices for high and low volatility days for different sectors. The heatmap represents the transition probabilities between the different types of orders as a grid of colored squares. Each cell’s color depicts the magnitude of the probability. The columns of the matrix represent the current state in the sequence whereas the rows represent the next state. Using these heatmaps, we analyze the dynamics of order transitions which is discussed below:
\begin{figure}
\centering
\begin{subfigure}{0.375\textwidth}
    \includegraphics[width=\textwidth]{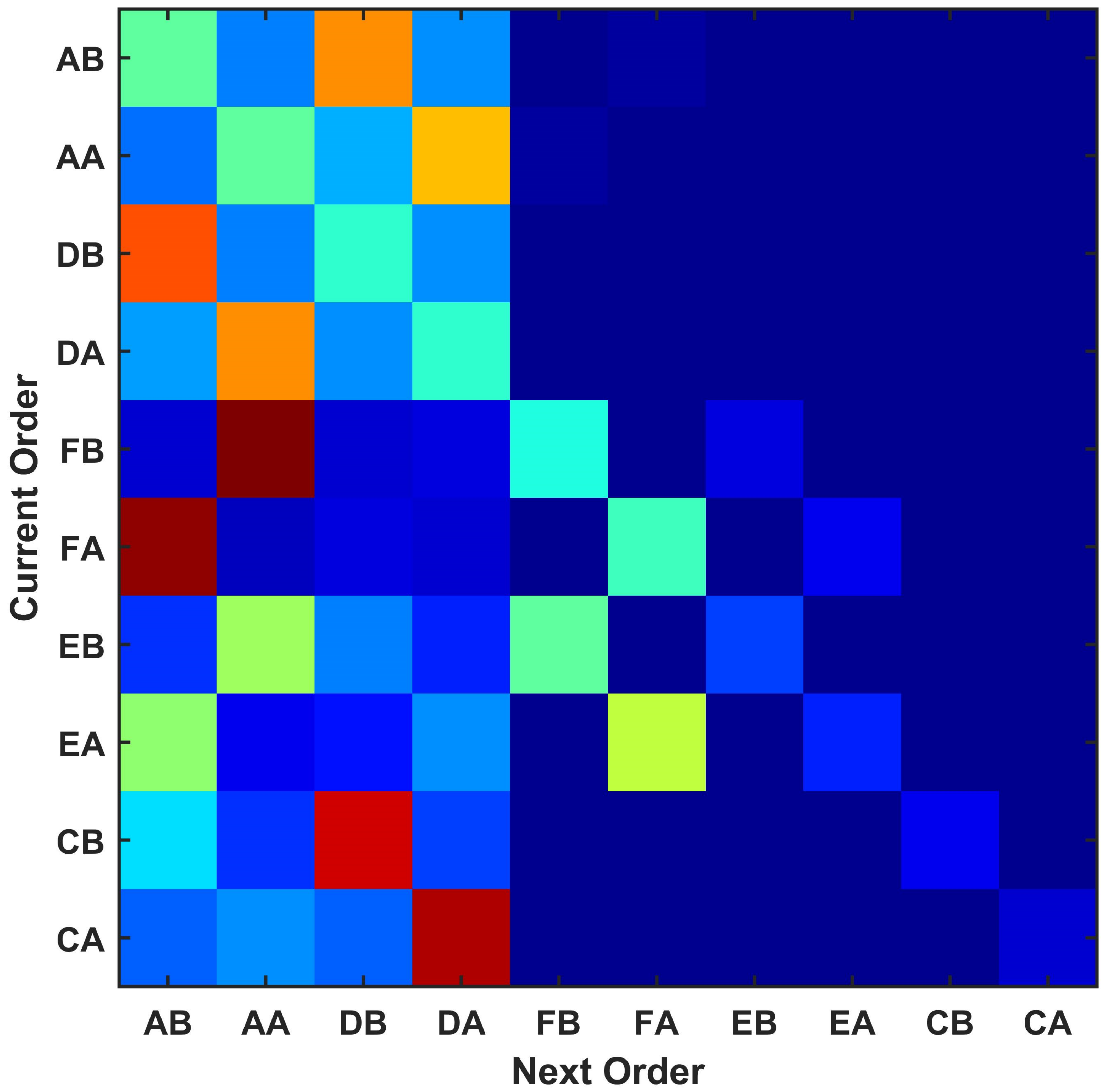}
    \caption{Energy (High volatility)}
    \label{subfig:E_H}
\end{subfigure}
\hspace{0.5cm}
\begin{subfigure}{0.405\textwidth}
    \includegraphics[width=\textwidth]{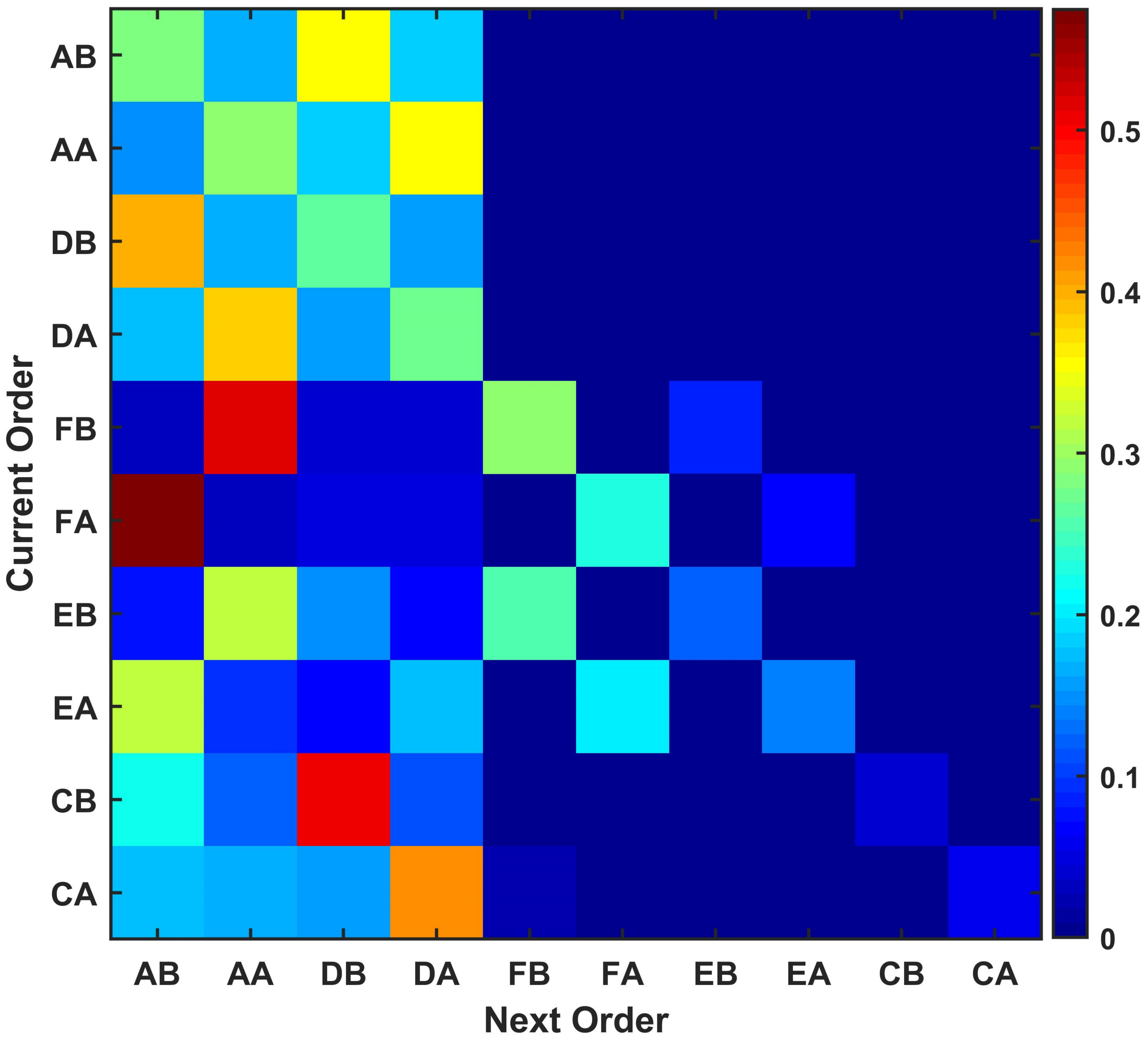}
    \caption{Energy (Low volatility)}
    \label{subfig:E_L}
\end{subfigure}
\begin{subfigure}{0.375\textwidth}
    \includegraphics[width=\textwidth]{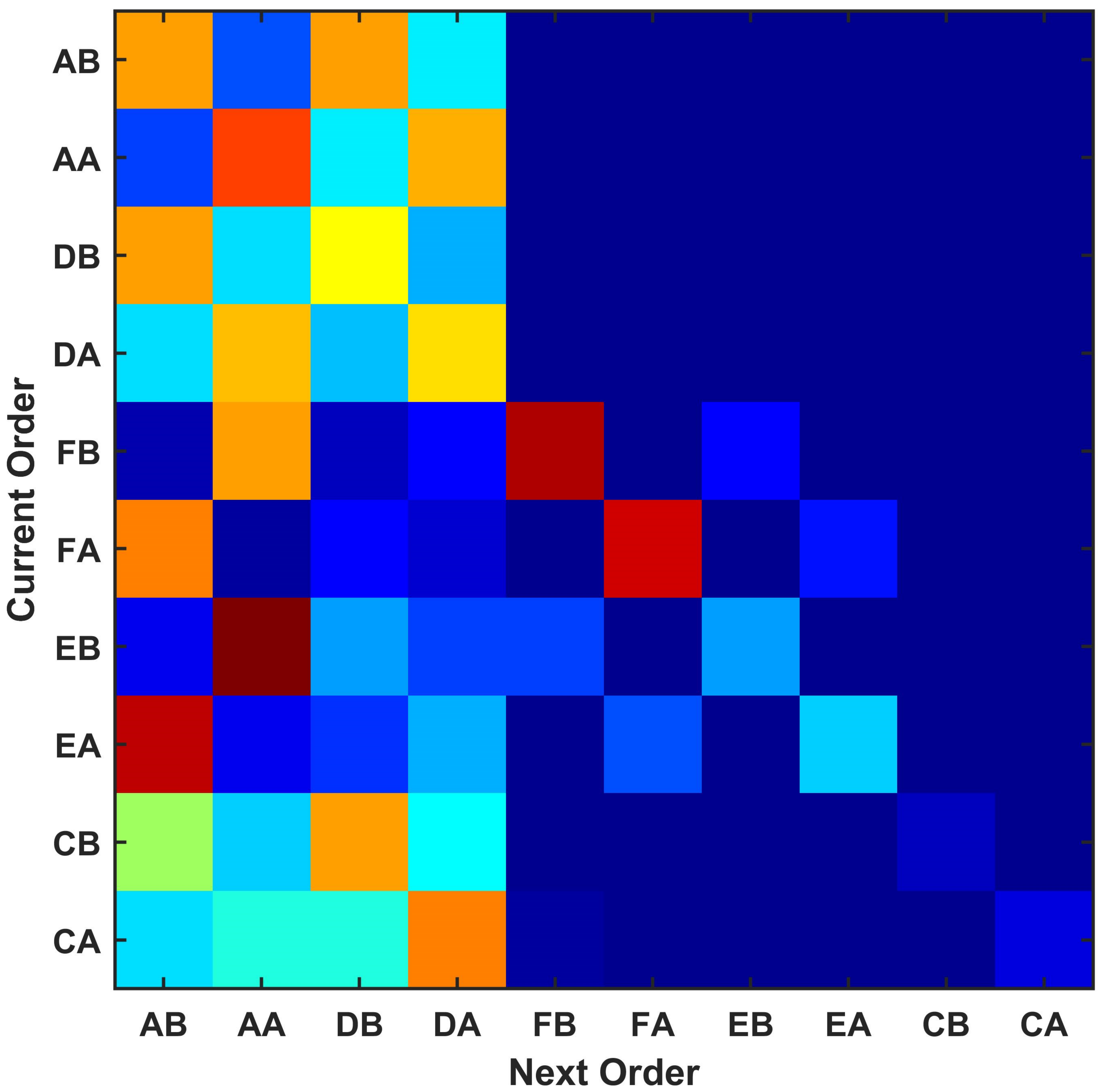}
    \caption{F\&B (High volatility)}
    \label{subfig:FB_H}
\end{subfigure}
\hspace{0.5cm}
\begin{subfigure}{0.405\textwidth}
    \includegraphics[width=\textwidth]{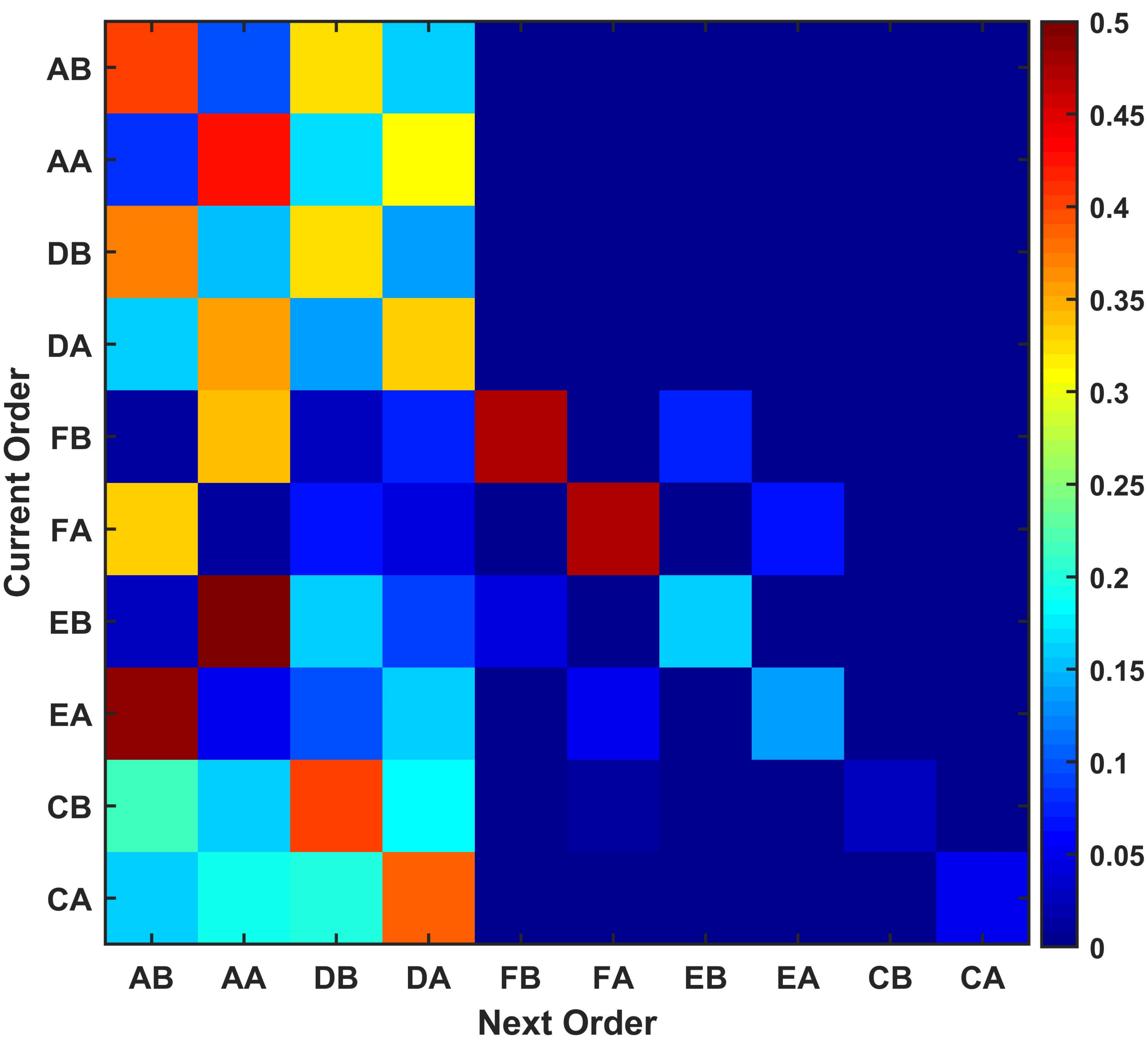}
    \caption{F\&B (Low volatility)}
    \label{subfig:FB_L}
\end{subfigure}
\begin{subfigure}{0.375\textwidth}
    \includegraphics[width=\textwidth]{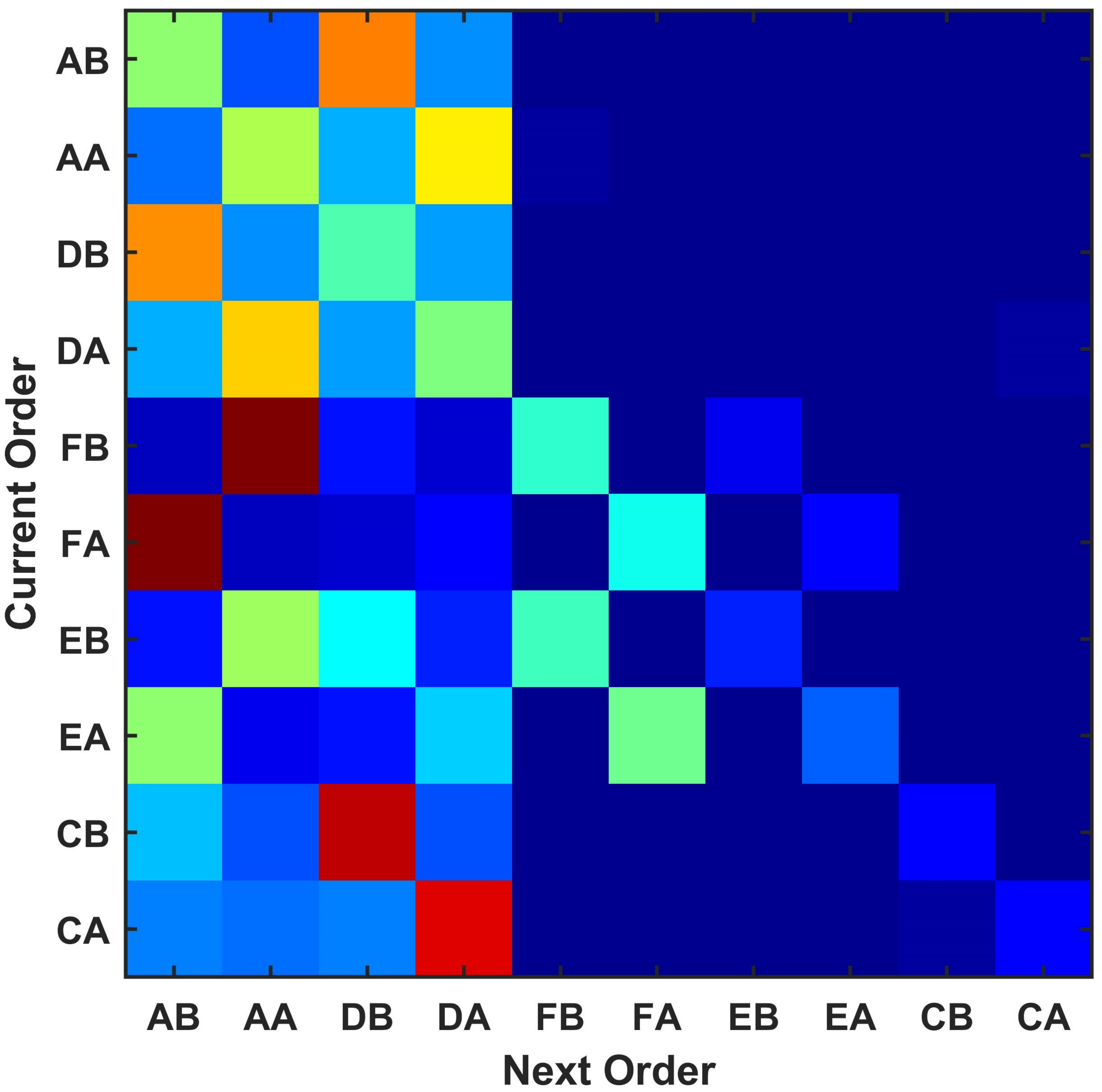}
    \caption{FMCG (High volatility)}
    \label{subfig:FMCG_H}
\end{subfigure}
\hspace{0.5cm}
\begin{subfigure}{0.405\textwidth}
    \includegraphics[width=\textwidth]{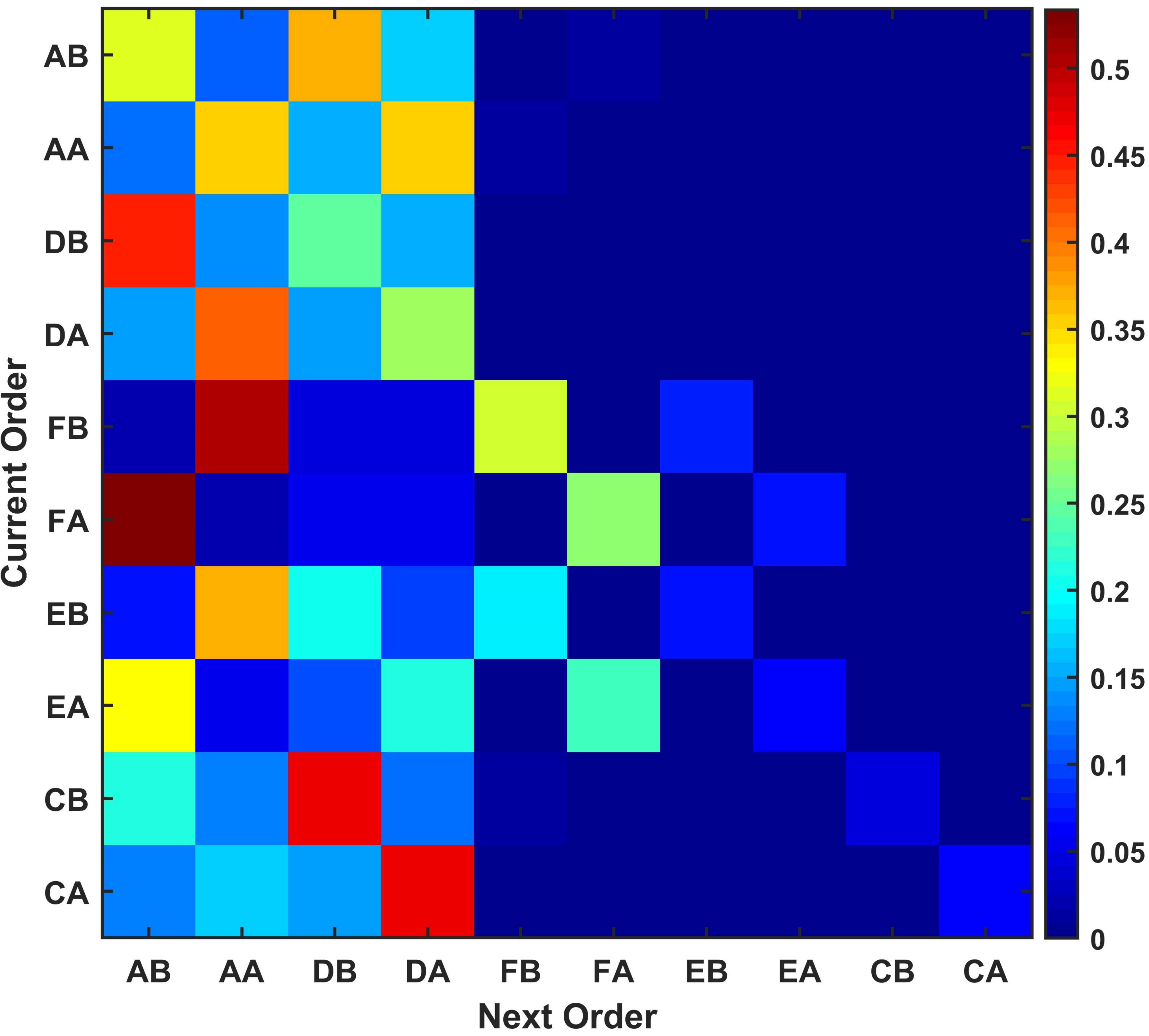}
    \caption{FMCG (Low volatility)}
    \label{subfig:FMCG_L}
\end{subfigure}      
\caption{Heatmap representation of the transition probability matrix on high and low volatility days.}
\label{fig:Heatmap1}
\end{figure}

\begin{enumerate}

\item The transition probability from $FB\rightarrow AA$ and $FA\rightarrow AB$ is greater on the high volatility days as compared to low volatility days for all the sectors. For $FB\rightarrow AA$, the average transition probability is $\approx 52\%$ on high volatility days compared to $\approx 47\%$ on low volatility days. This means that the probability of adding new buy limit orders right after a full execution of selling order is comparatively greater on high volatility days. Similarly, for $FA\rightarrow AB$, it is $\approx 51\%$ on high volatility days compared to $\approx 45\%$ on low volatility days. The high probability of adding limit order right after the full execution of an order indicates the presence of active traders who are continuously monitoring every movement of the market, especially on high volatility days. 

\item The probability of transitioning from an add order to a delete order is generally greater during the higher volatility days. On average, it is $\approx 39\%$ on high volatility days compared to $\approx 34\%$ on low volatility days. The high probability of placing delete orders right after adding limit orders could indicate that the traders are trying to manipulate the market by placing fake limit orders on high volatility days. The manipulator could create a false sense of demand by submitting limit orders that they immediately cancel. This can trick other traders into buying or selling at a price that is favorable to the manipulator.

\item We observe that the diagonal probability values of the heatmap matrix is less on the higher volatility days than the lower volatility days for all the sectors, as is evident from the Figures~\ref{fig:Heatmap1} and \ref{fig:Heatmap2}. This means that the degree of inertia~\cite{paxinou2021analyzing} of an order, i.e., the probability of an order to stay in the same order at the next time step, is generally less on high volatility days. It indicates the presence of different types of market participants during high volatility days; some experienced traders and some uninformed noise traders.

\end{enumerate}

It is also important to highlight that there is a very high probability of self-transition of full execution orders for the case of Finance \& Banking sector, as it is apparent from sub-figures~\ref{subfig:FB_H} and \ref{subfig:FB_L}. It means that there is a recurring pattern of $FB$ and $FA$. This is so because Finance \& Banking stocks were undervalued and less susceptible to uncertainties created by the trade war. Aided by the hike in interest rate by Federal Reserve, banking stocks were lucrative investment options at that moment. These assertions were supported by the positive results of banking stocks in the the last quarter (Q4) of the 2018 fiscal year. Therefore, investors looking to invest for the medium to long term, should look for sectors where there is minimum direct exposure to uncertainties if extreme market situations arise in the future. In future work, a detailed comparison between high-high and low-low volatility days of different sectors may be carried out.

\begin{table}[h]
\footnotesize
\setlength\tabcolsep{4pt}
\setcellgapes{2pt}
\makegapedcells
\renewcommand\multirowsetup{\centering} 

\begin{tabularx}{\linewidth}{ |p{5em}|l| *{12}{>{\centering\arraybackslash}X|}}
    \hline
    \textbf{Sector}&\textbf{Volatility}&\multicolumn{10}{c}{\textbf{Stationary distribution of the different orders}}\\
    \cline{3-12}   
    &      
        &   \thead{$\pi_{AB}$}  
            &   \thead{$\pi_{AA}$} 
                &   \thead{$\pi_{DB}$} 
                    &   \thead{$\pi_{DA}$} 
                        &   \thead{$\pi_{FB}$} 
                            &   \thead{$\pi_{FA}$} 
                                &   \thead{$\pi_{EB}$} 
                                    &   \thead{$\pi_{EA}$}
                                        &   \thead{$\pi_{CB}$}
                                            &   \thead{$\pi_{CA}$} \\ 
    \hline
\multirow{2}{=}{Energy} 
    &   High 
        &   0.256
            &   0.240
                &   0.245
                    &   0.231 
                         &   0.011
                             &   0.010
                                 &   0.003
                                    &   0.002 
                                        &   0.001     
                                             &  $<$0.001\\ 
                                      
    \cline{2-12}
    &   Low 
        &   0.249
            &   0.248
                &   0.239
                    &   0.241 
                         &   0.009
                             &   0.007
                                 &   0.003
                                    &   0.003 
                                        &   0.001
                                            &   0.001
                                            \\ 
    \hline
    \multirow{2}{=}{Finance\\Banking} 
    &   High 
        &   0.246
            &   0.250
                &   0.236
                    &   0.241
                         &   0.009
                             &   0.008
                                 &   0.003
                                    &   0.003 
                                        &   0.003     
                                            &   0.003
                                            \\ 
                                      
    \cline{2-12}
    &   Low 
        &   0.247
            &   0.250
                &   0.240
                    &   0.241 
                         &   0.006
                             &   0.006
                                 &   0.003
                                    &   0.002 
                                        &   0.002
                                            &   0.002
                                            \\ 
    \hline
    \multirow{2}{=}{FMCG} 
    &   High 
        &   0.257
            &   0.235
                &   0.247
                    &   0.227
                         &   0.010
                             &   0.009
                                 &   0.003
                                    &    0.003
                                        &   0.005     
                                            &   0.005
                                            \\ 
                                      
    \cline{2-12}
    &   Low 
        &   0.245
            &   0.252
                &   0.231
                    &   0.242 
                         &   0.012
                             &   0.009
                                 &   0.004
                                    &   0.003 
                                        &   0.001
                                            &   0.001
                                            \\ 
    \hline
    \multirow{2}{=}{Healthcare} 
    &   High 
        &   0.227
            &   0.268
                &   0.209
                    &   0.258 
                         &   0.014
                             &   0.015
                                 &   0.004
                                    &   0.004 
                                        &   $<$0.001     
                                            &   $<$0.001
                                            \\   
                                      
    \cline{2-12}
    &   Low 
        &   0.274
            &   0.221
                &   0.264
                    &   0.210 
                         &   0.012
                             &   0.010
                                 &   0.005
                                    &   0.005 
                                        &   $<$0.001
                                            &   $<$0.001
                                            \\ 
    \hline
    \multirow{2}{=}{IT} 
    &   High 
        &   0.258
            &   0.248
                &   0.236
                    &   0.226 
                         &   0.011
                             &   0.010
                                 &   0.003
                                    &   0.003 
                                        &   0.002 
                                            &   0.002
                                            \\ 
                                      
    \cline{2-12}
    &   Low 
        &   0.268
            &   0.239
                &   0.248
                    &   0.218 
                         &   0.008
                             &   0.009
                                 &   0.003
                                    &   0.004 
                                        &   0.001
                                            &   0.002
                                            \\ 
    \hline
    \multirow{2}{=}{Real-\\Estate} 
     &   High 
        &   0.278
            &   0.217
                &   0.267
                    &   0.203
                         &   0.013
                             &   0.013
                                 &   0.003
                                    &   0.004 
                                        &   0.001
                                            &   0.001
                                            \\ 
                                      
    \cline{2-12}
     &   Low 
        &   0.254
            &   0.237
                &   0.249
                    &   0.226 
                         &   0.009
                             &   0.009
                                 &   0.003
                                    &   0.005 
                                        &   0.001
                                            &   0.001
                                            \\ 
   
    \hline

\end{tabularx}
\caption {Stationary Distribution of the different types of orders for different sectors on high and low volatility days. The $\pi_{s}$ represents the stationary distribution probabilities and the subscripts to the $\pi_{s}$ represent the different types of orders.}
\label{tab:SD}
\end{table}

\subsection{Stationary Distribution}
\label{subsec:SD}
The stationary distribution of the DTMC refers to the long-run probability distribution that remains unchanged as time progresses. It  shows the long-term proportion of time each order spends in the sequence. The Markov chain for each stock representing the corresponding sectors is found to be ergodic and irreducible. As a result, the stationary distribution can be calculated for each of them. Table~\ref{tab:SD} shows the stationary distribution of sectors for the averaged high and low volatility days. 

We observe that $\pi_{AB}, \pi_{AA}, \pi_{DB}$ and $\pi_{DA}$ occupy more than $95\%$ of the total probability of all the orders during both high and low volatility days for every sector, which means that the highest proportion of time is spent on the transaction of these orders. In addition, execution orders, $\pi_{FB}, \pi_{FA}, \pi_{EB}, \pi_{EA}$ occupies less than $4\%$ of the total probability during both high and low volatility days, signifying that only a small proportion of the added limit orders are executed. These findings lead us to conclude that most of the limit orders are added with the intention to delete it, which is in line with the study~\cite{dalko2020high}. Such tactics are employed to create the appearance of liquidity coming into the market. The high probability of delete order and low probability of execution order indicates that the most of the traders intent to manipulate the market rather than executing trades. 

We also observe a slight difference in order dynamics between high and low-volatility days. $\pi_{FB}$ and $\pi_{FA}$ generally have comparatively higher values during high volatility days than the low volatility days. It means that the probability for full execution of buying and selling orders on high volatility days is high in the long run. The presence of more active participants during high volatility day could be the reason for high execution probability. 

\subsection{Mean Recurrence time (MRT)}
\label{subsec:MFPT_MRT}

Mean Recurrence time (MRT) is the expected number of time steps it takes for the chain to return to a particular state, starting from that state. The MRT of each state for the six sectors during high and low volatility days are calculated and shown in Table~\ref{tab:MRT}.

\begin{table}[ht!]
\footnotesize
\setlength\tabcolsep{5pt}
\setcellgapes{2pt}
\makegapedcells
\renewcommand\multirowsetup{\centering} % <-----

\begin{tabularx}{\linewidth}{ |p{5em}|c| *{10}{>{\centering\arraybackslash}X|} }
    \hline
\thead{Sector} 
    &   \thead{Volatility} 
        &   \thead{AB}  
            &   \thead{AA} 
                &   \thead{DB} 
                    &   \thead{DA} 
                        &   \thead{FB} 
                            &   \thead{FA} 
                                &   \thead{EB} 
                                    &   \thead{EA}
                                        &   \thead{CB}
                                            &   \thead{CA} \\ 
    \hline
\multirow{2}{=}{Energy} 
    &   High 
        &   3.9
            &   4.2
                &   4.1
                    &   4.3 
                         &   93.2
                             &   103.2
                                 &   358.5
                                    &   421.6 
                                        &   772.4     
                                             &   1002.1\\ 
                                      
    \cline{2-12}
    &   Low 
        &   4.0
            &   4.0
                &   4.2
                    &   4.1 
                         &   109.0
                             &   144.1
                                 &   315.7
                                    &   332.9 
                                        &   1364.2
                                            &   1116.7
                                            \\ 
    \hline
    \multirow{2}{=}{Finance\\Banking} 
    &   High 
        &   4.1
            &   4.0
                &   4.2
                    &   4.2
                         &   115.0
                             &   125.2
                                 &   375.7
                                    &   382.3 
                                        &   339.3     
                                            &   371.9
                                            \\ 
                                      
    \cline{2-12}
    &   Low 
        &   4.0
            &   4.0
                &   4.2
                    &   4.1 
                         &   167.5
                             &   179.7
                                 &   382.1
                                    &   400.1 
                                        &   476.7
                                            &   436.4
                                            \\ 
    \hline
    \multirow{2}{=}{FMCG} 
    &   High 
        &   3.9
            &   4.2
                &   4.1
                    &   4.4 
                         &   97.2
                             &   112.4
                                 &   392.2
                                    &    398.4
                                        &   200.6     
                                            &   187.9
                                            \\ 
                                      
    \cline{2-12}
    &   Low 
        &   4.1
            &   4.0
                &   4.3
                    &   4.1 
                         &   85.5
                             &   116.8
                                 &   263.5
                                    &   350.9 
                                        &   899.5
                                            &   844.2
                                            \\ 
    \hline
    \multirow{2}{=}{Healthcare} 
    &   High 
        &   4.4
            &   3.7
                &   4.8
                    &   3.9 
                         &   69.7
                             &   66.9
                                 &   246.7
                                    &   262.2 
                                        &   3405.3     
                                            &   3084.7
                                            \\   
                                      
    \cline{2-12}
    &   Low 
        &   3.7
            &   4.5
                &   3.8
                    &   4.8 
                         &   80.9
                             &   102.6
                                 &   206.6
                                    &   206.0 
                                        &   3481.5
                                            &   2160.4
                                            \\ 
    \hline
    \multirow{2}{=}{IT} 
    &   High 
        &   3.9
            &   4.0
                &   4.2
                    &   4.4 
                         &   87.9
                             &   95.6
                                 &   310.5
                                    &   309.3 
                                        &   531.2 
                                            &   623.6
                                            \\ 
                                      
    \cline{2-12}
    &   Low 
        &   3.7
            &   4.2
                &   4.0
                    &   4.6 
                         &   121.6
                             &   116.0
                                 &   329.0
                                    &   283.6 
                                        &   1366.0
                                            &   409.0
                                            \\ 
    \hline
    \multirow{2}{=}{Real-\\Estate} 
     &   High 
        &   3.6
            &   4.6
                &   3.7
                    &   4.9 
                         &   79.7
                             &   78.6
                                 &   328.5
                                    &   273.0 
                                        &   1438.1
                                            &   866.7
                                            \\ 
                                      
    \cline{2-12}
     &   Low 
        &   3.9
            &   4.2
                &   4.0
                    &   4.4 
                         &   110.6
                             &   106.6
                                 &   293.9
                                    &   217.6 
                                        &   988.2
                                            &   700.2
                                            \\ 
   
    \hline

\end{tabularx}
\caption {Mean Recurrence time of the different types of orders for different sectors at high and low volatility days. The subscripts represent the different types of orders.}
\label{tab:MRT}
\end{table}

From Table~\ref{tab:MRT}, we observe that MRT values for the orders $AB$, $AA$ and $DB$, $DA$ are significantly less than the rest of the orders. The MRT of $AB$ and $AA$ is closely followed by the MRT of $DB$ and $DA$. This shows that the traders are mostly adding and deleting limit orders, which indicates that most of the submitted orders are fake. They are placed to manipulate the uninformed traders and the market in general, as shown in~\cite{dalko2020high}. On comparing the MRT during high and low volatility days, we observe that the MRT values of $FB$ and $FA$ are less during high volatility day than the low volatility day. The presence of more active traders during high volatility days may be the reason for low MRT value for $FB$ and $FA$.

We have highlighted about the high self-transition probability of the orders, $FB$ and $FA$ for Finance \& Banking sector where there is a recurring pattern in Section~\ref{subsec:TPM}. Here, we notice that the MRT of $FB$ and $FA$ for Finance \& Banking is relatively high in comparison to other sectors. It means that the recurring pattern of $FB$ and $FA$ occurs very far apart. This may indicate that uninformed market participants wait and watch the price movement of banking stocks. As soon as they observed orders being executed, presumably by experienced and informed traders, they also start placing orders at market price so that their orders are also executed. Following the trading action of informed traders by uninformed ones is understandable as there were lot of fears and uncertainties in the market during the trade war.

The mean recurrence time of stock market orders can provide insights into various aspects of market activity, including order flow and trader behavior. It is a valuable metric for traders, market analysts, and researchers looking to understand and respond to changing market conditions and trading patterns.

\subsection{Spectral Gap, Relaxation Rate and Entropy Rate}
\label{subsec:SG_RR_ER}
We have seen that an ergodic and irreducible DTMC on a finite state space $S$ converges to a unique stationary distribution $\pi$. Besides these convergence, we will also enquire about the rate of convergence. The spectral gap and relaxation rate provide different perspectives on the convergence rate and the behavior of the Markov chain. Whereas, the entropy measures the amount of randomness or unpredictability in the sequence of states.

\begin{figure}[ht!]
\centering
\begin{subfigure}{0.4\textwidth}
    \includegraphics[width=\textwidth]{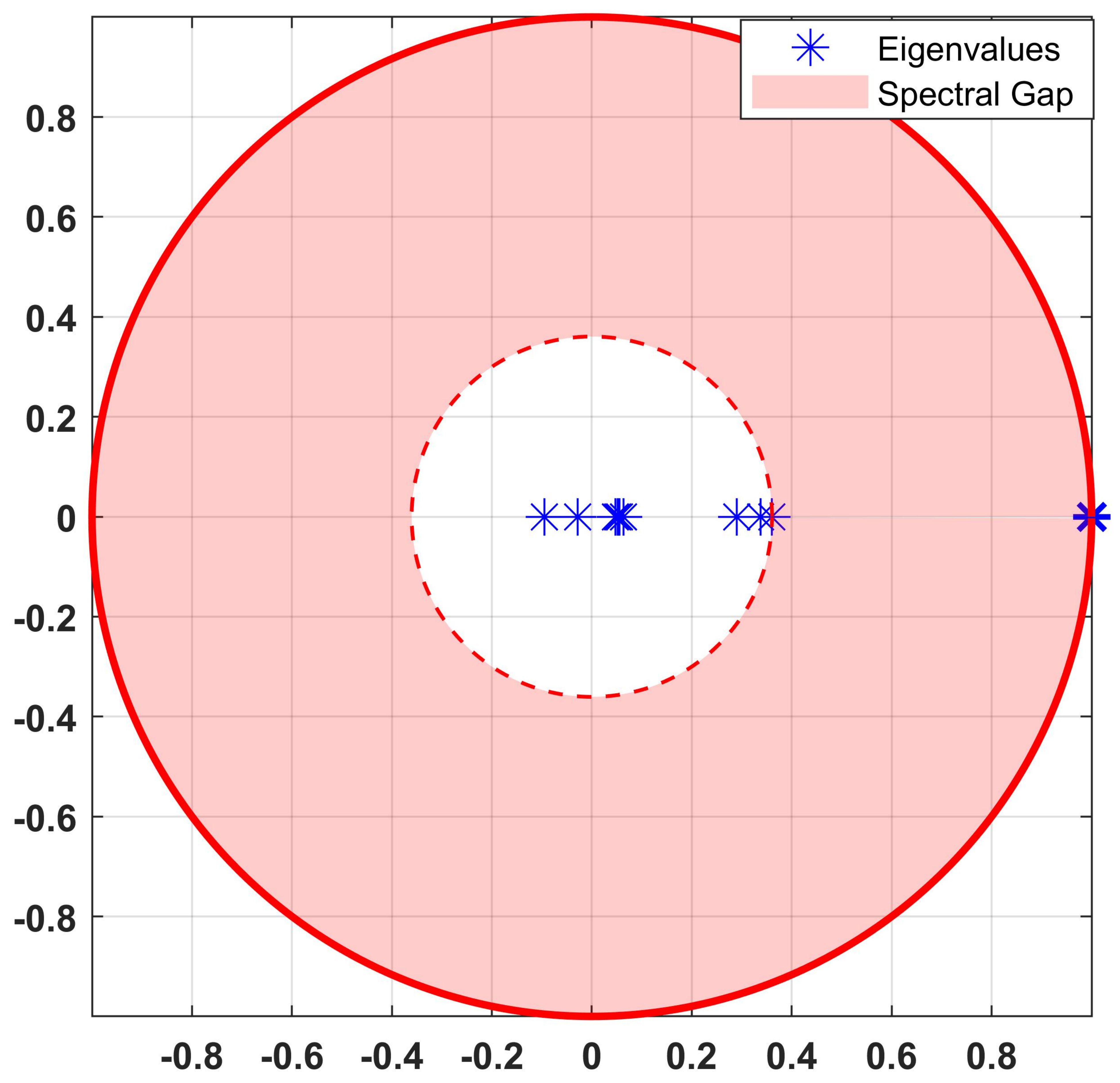}
    \caption{Energy(High volatility)}
    \label{subfig:SG1}
\end{subfigure}
\hspace{1cm}
\begin{subfigure}{0.4\textwidth}
    \includegraphics[width=\textwidth]{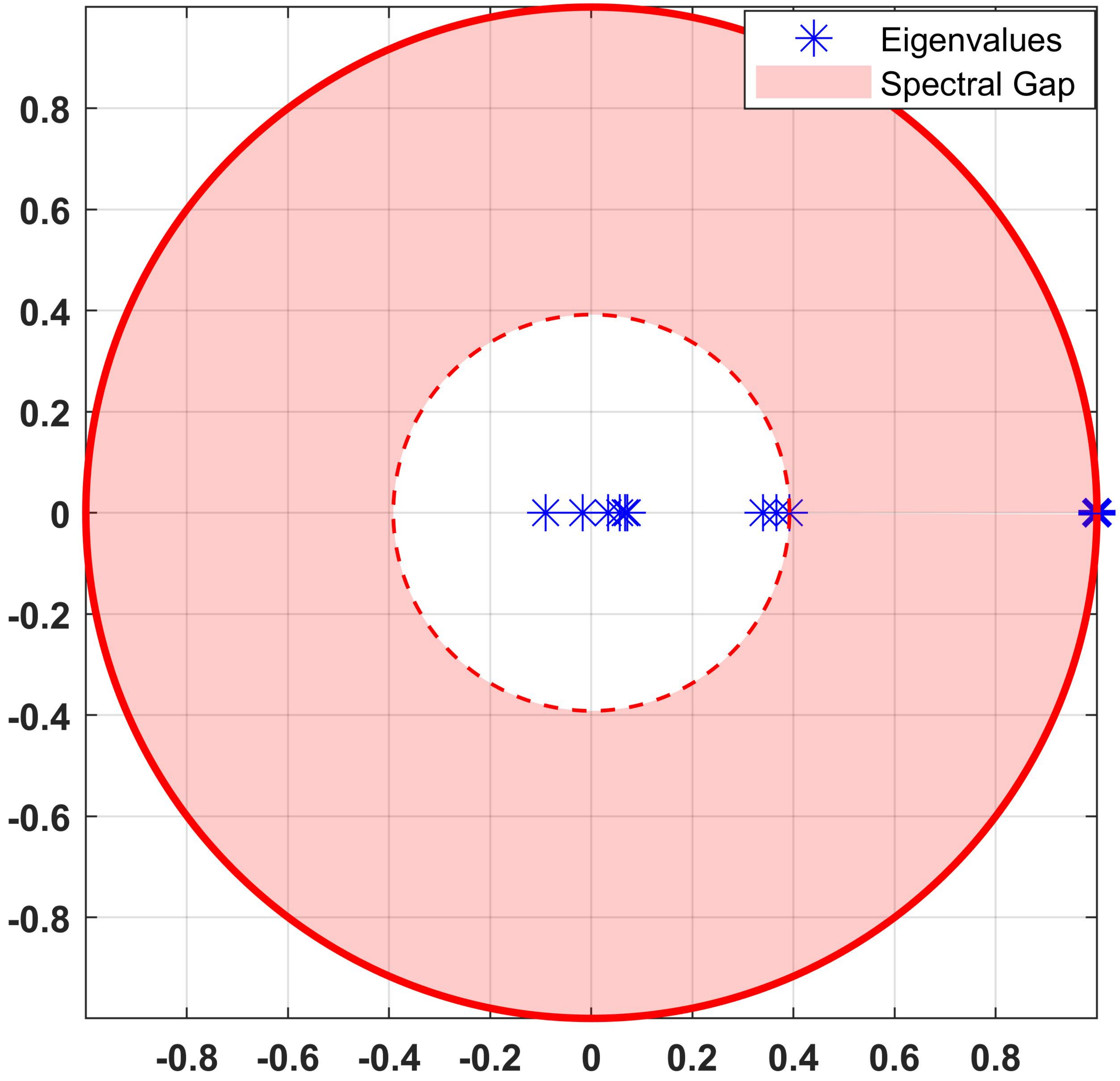}
    \caption{Energy(Low volatility)}
    \label{subfig:SG2}
\end{subfigure}
       
\caption{Spectral Gap of the DTMC for IT sector.}
\label{fig:SG}
\end{figure}

The spectral gap of the DTMC is the difference between the two largest eigenvalues of the transition probability matrix of the chain and the relaxation rate is the inverse of the absolute value of the second largest eigenvalue of the matrix. The spectral gap and the relaxation rate determines the convergence speed of the chain to its stationary distribution. Figure~\ref{fig:SG} show the eigenvalues on the complex plane and spectral gap for IT sector. The blue stars represents the eigenvalues and the shaded red area gives the spectral gap of the chain. 

From Table~\ref{tab:ER}, we notice that the difference of the spectral gap and relaxation rate between high and low volatility days is very small and insignificant. This small difference indicates that the rate of convergence to the long-term behaviour of different types of orders on high and low volatility days are similar. This indicates that orders transition to its long-term probability, as given in Table~\ref{tab:SD}, at a similar rate. It may imply that the traders employ similar trading strategies or objectives on both on high and low volatility days during the trade war. Further, in the case of Finance \& Banking sector, we observe a small spectral gap and a large relaxation rate compared to other sectors. The small gap indicates that the chain retains memory of previous states for a longer duration. This long-term memory delays the convergence to the steady states. 

\begin{table}[!h]
\footnotesize
   \centering
    \begin{tabular}{ |p{2.0cm}|p{2.0cm}|p{1.5cm}|p{1.5cm}|p{1.5cm}|p{1.5cm}|p{1.5cm}|p{1.5cm}| }
 
 \hline
\thead{Properties} 
    &   \thead{Volatility} 
        &   \thead{Energy}  
            &   \thead{Finance\\Banking} 
                &   \thead{FMCG} 
                    &   \thead{Health-\\care} 
                        &   \thead{IT} 
                            &   \thead{Real\\Estate} 
                                 \\ 
    \hline
    \multirow{2}{=}{Spectral Gap} 
    &   High 
        &   0.614 
            &   0.508
                &   0.604
                    &   0.608
                         &   0.640
                             &   0.586
                                 \\ 
                                      
    \cline{2-8}
    &   Low 
        &   0.613 
            &   0.494
                &   0.577
                    &   0.598 
                         &   0.608
                             &   0.614
                                 \\ 
    \hline
    \multirow{2}{=}{Relaxation Rate} 
    &   High 
        &   1.618 
            &   1.969
                &   1.656
                    &   1.645 
                         &   1.563
                             &   1.706
                                 \\ 
                                      
    \cline{2-8}
    &   Low 
        &   1.631 
            &   2.024
                &   1.733
                    &   1.672 
                         &   1.645
                             &   1.629
                                 \\ 
   
    \hline
    \multirow{2}{=}{Entropy Rate} 
    &   High 
        &   2.006 
            &   2.001
                &   2.060
                    &   1.999 
                         &   2.079
                             &   1.991
                                 \\ 
                                      
    \cline{2-8}
    &   Low 
        &   2.024 
            &   1.947
                &   2.001
                    &   1.952 
                         &   2.025
                             &   2.030
                                 \\ 
    \hline
   
\end{tabular}
\caption{Properties, namely, Spectral Gap, Relaxation Rate and Entropy Rate of the DTMC model for high and low volatility days of different sectors.}
\label{tab:ER}
\end{table}

The entropy rate of a DTMC is the expected amount of uncertainty about the next order of the sequence, given a current order. The maximum entropy rate of the order sequence is $log_2 r = 3.322$. From Table~\ref{tab:ER}, we observe that the entropy rates for high and low volatility days are similar. There is around 60\% uncertainties on both volatility days. It implies that the orders in the sequence exhibit a similar level of uncertainty or randomness in their transitions.  The distribution of orders at each time step is almost similar. This signifies that market sentiments of the various market participants are similar during the trade war.

The spectral gap can provide insights about how fast or slow the stock market orders are changing. The relaxation rate gives the rate of convergence to stationary distribution which is indicative of either a stress market or a calm market. The entropy rate gives information about the predictability in the placement of orders. Traders may incorporate these insights into their trading and investment decisions to capitalize on the market conditions and optimize their outcomes.

\section{Conclusions}
\label{sec:Conc}
In this study, we used a first-order time-homogeneous discrete time Markov chain (DTMC) model to analyze the high-frequency stock market order transition dynamics of an order sequence during the USA-China trade war of 2018. Our findings reveal that the probability of placing limit orders immediately following the complete execution of an order, i.e., $FB\rightarrow AA$ and $FA\rightarrow AB$, are high on high volatility days compared to low volatility days. It indicates the presence of active traders who are continuously monitoring every movement of the market on such days. Further, the low degree of inertia of orders during high-volatility days shows the presence of different types of active market participants on such days. We also found that the transition probability from an add orders to a delete orders, i.e., $AB\rightarrow DB$ and $AA\rightarrow DA$, are very high, which suggests that traders place limit orders primarily with the intention of deleting the majority of them to influence the market on high volatility days. The result also shows a recurring pattern of the full execution of bid and ask orders, i.e., $FB$ and $FA$, in the order sequence of banking stocks. Small spectral gap and entropy rate and a large relaxation rate of Finance \& Banking sector confirms recurring the pattern. A high mean recurrence time (MRT) for $FB$ and $FA$ also indicates that full order execution occurs very far apart. It signifies that uninformed market participants wait and watch the price movement of stocks belonging to Finance \& Banking sector and follow the trading action of informed traders in order to execute such stocks during the trade war.

High volatility days during uncertain market periods can be a good time to trade as there is a lot of price movement, but it is important to remember that there is a greater risk of losing money as market manipulators are also looking to take advantage during such times. This study can help understand and adapt to market dynamics and assess risk in response to varying levels of volatilities during extreme macroeconomic periods. Additionally, market participants should look out for sectors which are undervalued and resilient to uncertainties during extreme macroeconomic conditions such as Finance \& Banking sector.

The first-order time-homogeneous DTMC analysis conducted in this study can be extended to include other periods of extreme market conditions, such as the financial crisis of 2008 and the Covid-19 pandemic. By broadening the analysis to include these periods, we can enhance our understanding of how extreme market conditions influence trading behavior. Additionally, exploring order transition dynamics during pre-market hours, market hours, and after-market hours, while considering the size of each order, could yield valuable insights for future studies. Furthermore, a thorough comparison between high-high and low-low volatility days across different sectors could enhance our comprehension of order dynamics.

\section{Acknowledgement}

We extend our gratitude to Algoseek~\cite{Algo} for graciously providing the data necessary for our analysis at free of cost. We also like to appreciate the help of Kundan Mukhia and Shubham Priyadarshi. Additionally, we would like to express our appreciation to the Director of National Institute of Technology, Sikkim for awarding a PhD scholarship to Salam Rabindrajit Luwang and Anish Rai.

\section{Data Availability}

The data for this study were provided by Algoseek~\cite{Algo} through a legal agreement at free of cost. 

\bibliographystyle{elsarticle-num} 
\bibliography{References}

\newpage

\appendix
\section{Data Description}

\begin{table}[ht!]
\large
\begin{center}
\begin{adjustbox}{width=\columnwidth,center}
\begin{tabular}{|c|c|c|l|l|c|c|l|}
  \hline
  DATE & TIMESTAMP & ORDER ID. & EVENT TYPE & TICKER & PRICE & QUANTITY & EXCHANGE \\
  \hline
   2018-11-06 & 4:00:00.002 & 12011 & ADD-BID & AAPL & 164.99 & 100 & NASDAQ \\
   \hline
    2018-11-06 & 4:00:00.032 & 12056 & ADD-ASK & AAPL & 194.99 & 500 & NASDAQ \\
    \hline
    2018-11-06 & 4:00:00.112 & 13473 & ADD-BID & XLF  &  67.50 & 300 & NASDAQ \\
    \hline
    ... & ... & ... & ... & ... & ... & ... & ... \\ 
    \hline
    ... & ... & ... & ... & ... & ... & ... & ... \\ 
    \hline
    2018-11-06 & 9:30:00.156 & 89017 & DELETE-BID & GOOGL & 0 & 100 & NASDAQ \\ 
    \hline
    2018-11-06 & 9:30:01.006 & 83907 & ADD-BID & INTC & 123.70 & 200 & NASDAQ \\
    \hline
    ... & ... & ... & ... & ... & ... & ... & ... \\ 
    \hline
    ... & ... & ... & ... & ... & ... & ... & ... \\ 
    \hline
    2018-11-06 & 16:00.00.000 & 123483 & DELETE-BID & AMD & 0 & 150 & NASDAQ \\ 
    \hline
    ... & ... & ... & ... & ... & ... & ... & ... \\ 
    \hline
    ... & ... & ... & ... & ... & ... & ... & ... \\ 
    \hline
    2018-11-06 & 20.00.00.000 & 547324 & DELETE-ASK & NVDA & 0 & 40 & NASDAQ \\
    \hline
\end{tabular}
\end{adjustbox}
\caption{Table represents a sample format of the high-frequency data. The columns represents the trading day, the Timestamp (in milliseconds) of placing the order, order ID, Event type, Ticker, Price and Quantity and Exchange.}
\label{tab:DataSample}
\end{center}
\end{table}

\begin{table}[!h]
\begin{center}
\small
    \begin{tabular}{ | l | l | p{6.5cm} |}
    \hline
	  Order (State) & Abbreviation & Description \\
	\hline
ADD-BID & AB & Add Bid order  \\
\hline
ADD-ASK & AA & Add Ask order  \\
\hline
CANCEL-BID & CB & Cancel outstanding Bid order in part  \\
\hline
CANCEL-ASK & CA & Cancel outstanding Ask order in part  \\
\hline
DELETE-BID & DB & Delete outstanding Bid order in full  \\
\hline
DELETE-ASK & DA &Delete outstanding Ask order in full  \\
\hline
EXECUTE-BID & EB & Execute outstanding Bid order in part  \\
\hline
EXECUTE-ASK & EA & Execute outstanding Ask order in part  \\
\hline
FILL-BID & FB & Execute outstanding Bid order in full  \\
\hline
FILL-ASK & FA & Execute outstanding Ask order in full  \\
\hline
    \end{tabular}
    \caption{Names and descriptions of the type of stock market orders found in the dataset.}
    \label{tab:EventDescription}
\end{center}
\end{table}

\section{Results}
\subsection{Selection of High and Low volatility days}
\begin{table}[ht!]
\footnotesize
\begin{center}
\begin{tabular}{|l|l|l|l|l|l|}
   \hline
   \textbf{Energy} & \textbf{Finance \& Banking} & \textbf{FMCG} & \textbf{Healthcare} & \textbf{IT} & \textbf{Real estate} \\
   \hline
   NextEra Energy  & Bank of America & Johnson \&  & UnitedHealth  & Apple  & American Tower  \\
   Inc(NEE) &  Corp(BAC) &  Johnson(JNJ) & Group Inc(UNH) &  Inc(AAPL) &  Corp(AMT) \\
   \hline
   Exxon Mobil  & JPMorgan Chase  & McDonald's  & CVS Health& Microsoft  & Crown Castle   \\
    Corp(XOM) & \& Co(JPM) &Corp(MCD) &Corp(CVS) & Corp(MSFT) &  Inc(CCI)  \\
   \hline
   Chevron & Wells Fargo  & Procter \&  & McKesson & Alphabet  & Prologis\\
   Corporation(CVX) & \& Co(WFC) & Gamble Co(PG) & Corp(MCK) & Inc(GOOGL) &Inc(PLD) \\
   \hline   
\end{tabular}
\caption{Sector-wise composition of stocks for this study.}
\label{tab:Stocks_Sectors}
\end{center}
\end{table}

\begin{table}[ht!] 
\tiny
\setlength\tabcolsep{2pt}
\setcellgapes{2pt}
\makegapedcells
\renewcommand\multirowsetup{\centering}
\begin{tabularx}{\linewidth}{ |p{5em}|c|X|X|X| }
    \hline
\thead{Sector} 
    &   \thead{Volatility} 
        &   \thead{Company} 
            &   \thead{$\chi^2$-Statistic} 
                &   \thead{p-value} \\ 
    \hline
\multirow{6}{=}{Energy} 
    &    
        &   CVX
            &   137603.36
                &   $<$0.001 \\
    &   High 
        &   NEE
            &   86927.86
                &   $<$0.001 \\
    &   
        &   XOM
            &   431199.07
                &   $<$0.001 \\ 
    \cline{2-5}
    &    
        &   CVX
            &   103960.45
                &   $<$0.001 \\
    &   Low 
        &   NEE
            &   95142.45
                &   $<$0.001 \\
    &    
        &   XOM
            &   560936.75
                &   $<$0.001 \\
    \hline
\multirow{6}{=}{Finance\\Banking} 
    &    
        &   BAC
            &   1669922.36
                &   $<$0.001 \\
    &   High 
        &   JPM
            &   556042.84
                &   $<$0.001 \\
    &    
        &   WFC
            &   876874.03
                &   $<$0.001 \\
    \cline{2-5}
    &    
        &   BAC
            &   1616643.68
                &   $<$0.001 \\
    &   Low 
        &   JPM
            &   416412.91
                &   $<$0.001 \\
    &   
        &   WFC
            &   518309.52
                &   $<$0.001 \\
    \hline
\multirow{6}{=}{FMCG} 
    &    
        &   JNJ
            &   107319.61
                &   $<$0.001 \\
    &   High 
        &   MCD
            &   138118.21
                &   $<$0.001 \\
    &    
        &   PG
            &   272398.41
                &   $<$0.001 \\
    \cline{2-5}
    &    
        &   JNJ
            &   145333.49
                &   $<$0.001 \\
    &   Low 
        &   MCD
            &   103153.92
                &   $<$0.001 \\
    &    
        &   PG
            &   238712.92
                &   $<$0.001 \\
    \hline
\multirow{6}{=}{Healthcare} 
    &    
        &   CVS
            &   196551.34
                &   $<$0.001 \\
    &   High 
        &   MCK
            &   83699.06
                &   $<$0.001 \\
    &    
        &   UNH
            &   112895.92
                &   $<$0.001 \\
    \cline{2-5}
    &    
        &   CVS
            &   215840.32
                &   $<$0.001 \\
    &   Low 
        &   MCK
            &   39379.17
                &   $<$0.001 \\
    &    
        &   UNH
            &   119949.49
                &   $<$0.001 \\
    \hline
\multirow{6}{=}{IT} 
    &    
        &   AAPL
            &   573574.96
                &   $<$0.001 \\
    &   High 
        &   GOOGL
            &   1381442.81
                &   $<$0.001 \\
    &    
        &   MSFT
            &   1030025.57
                &   $<$0.001 \\
    \cline{2-5}
    &   
        &   AAPL
            &   497352.15
                &   $<$0.001 \\
    &   Low 
        &   GOOGL
            &   718397.31
                &   $<$0.001 \\
    &    
        &   MSFT
            &   1381752.48
                &   $<$0.001 \\
    \hline
\multirow{6}{=}{Real-Estate} 
    &    
        &   AMT
            &   69873.06
                &   $<$0.001 \\
    &   High 
        &   CCI
            &   115779.40
                &   $<$0.001 \\
    &    
        &   PLD
            &   85339.61
                &   $<$0.001 \\
    \cline{2-5}
    &    
        &   AMT
            &   66474.84
                &   $<$0.001 \\
    &   Low 
        &   CCI
            &   54315.35
                &   $<$0.001 \\
    &    
        &   PLD
            &   127174.90
                &   $<$0.001 \\
    \hline
\end{tabularx}
\caption{$\chi^2$-test to check the dependence of orders in the sequence of high-frequency order transactions. The test affirms the feasibility of applying Markov property to the order sequence.}
\label{tab:Chi}
\end{table}

\end{document}